\documentclass[aps, 11pt,a4paper,twocolumn,groupedaddress,preprintnumbers,nofootinbib,floatfix]{revtex4-1}
\usepackage[top=2.9cm, bottom=2.1cm, left=2cm, right=2cm]{geometry}

\usepackage[english]{babel}
\usepackage{amsmath}
\usepackage{graphicx}
\usepackage{color}
\usepackage{amssymb}
\usepackage{hyperref}
\usepackage{subfigure}
\usepackage{dcolumn}
\usepackage{bm}
\usepackage{graphicx}
\usepackage{float}

\newcommand{\meff}{\mbox{$\left|  m_{\beta\beta} \right|$ }}
\newcommand{\sumass}{\mbox{$\sum_i m_i$ }}
\newcommand{\betabeta}{\mbox{$(\beta \beta)_{0 \nu}  $}}

\begin{document}

\preprint{RM3-TH/14-11}

 \title{The quest for neutrinoless double beta decay: \\Pseudo-Dirac, Majorana and sterile neutrinos}

\author{
A. Meroni,$^{1,2}$\footnote{Electronic address:ameroni@fis.uniroma3.it}
E. Peinado,$^{2}$\footnote{Electronic address:epeinado@lnf.infn.it}}
\affiliation{
$^1$Dipartimento di Matematica e Fisica, Universit\`a di Roma Tre, Via della Vasca Navale 84, I-00146 Rome, Italy\\
$^2$ INFN, Laboratori Nazionali di Frascati, Via Enrico Fermi 40, I-00044 Frascati, Italy}

\begin{abstract}
        In this paper we analyze the neutrinoless double beta decay predictions 
        in some scenarios with admixture of pseudo-Dirac and  Majorana neutrinos  
        in the 3 and 3+1 neutrino frameworks. 
        We found that some of the cases can be falsifiable in  near-term and future 
        generations of neutrinoless double beta decay experiments even for the normal 
        neutrino mass hierarchy. In the 3+1 framework we consider the sterile neutrino 
        with a mass of the order of 1 $eV$. The complementarity between cosmological 
        constraints and the future sensitivity  for the next generations of the 
        neutrinoless double beta decay searches is exploited.  
\end{abstract}
\maketitle

\section{Introduction}

         The observation of neutrino oscillations \cite{art:2012,An:2012eh, Ahn:2012nd, Abe:2011sj} 
         implies that neutrinos are massive particles. 
         The fact that the  neutrino masses scale is several orders of magnitude smaller than 
         the rest of the fermions of the Standard Model (SM)\footnote{The most recent cosmological 
         upper bound for the sum of neutrino masses is $\sum m_\nu \le (0.2-0.6) eV$~\cite{Ade:2013zuv},
         this will be complemented by the future sensitivity from the Katrin 
         experiment~\cite{Bornschein:2003xi,fortheKATRIN:2013saa}.}, suggests
         an extension in which neutrinos are generally 
         expected to be of Majorana type, violating the total lepton number symmetry,
	 such as for example the so-called seesaw mechanism~\cite{Minkowski:1977sc,Yanagida:1979as,GellMann:1980vs,Schechter:1980gr,Mohapatra:1980yp} 
	 which accounts for the observed smallness of neutrino mass relative to that 
	 of charged fermions. 

The only feasible underground experiments at present that could be able to pin down the Majorana nature of
massive neutrinos, namely to prove the electron neutrino Majorana effective mass $\meff$,  and thus give information on
the violation of the total lepton number  symmetry~\cite{valle:1983yw} are   those  
searching for neutrinoless double beta (\betabeta)-decay \cite{Racah:1937qq}: $(A,Z) \rightarrow (A,Z+2) + e^- + e^-$.

Despite intense ongoing efforts, the decay has not been observed yet,
but important progresses have been made and especially the data from the GERDA-I
\cite{Agostini:2013mba} have shown that the claim by Klapdor-Kleingrothaus {\it et al.}~\cite{01-Klap04} is now strongly disfavored (see \cite{Barabash:2011fn,Piquemal:2013uaa}
for experimental review). On the other hand $|\Delta L|= 2$ processes can be tested in colliders, which opens the possibility to investigate all the elements of the neutrino mass matrix, $M_{\alpha \beta}^\nu$ with $\alpha,\beta= e, \mu,\tau$
(for details see \cite{hirsch:2006yk}). 

In summary the status of lepton and baryon number symmetries remains one 
of the most interesting unsolved questions in particle physics~\cite{weinberg:1980bf}, 
and neutrinos could very well be Dirac fermions.
\\
\section{Neutrinoless double beta decay: General Framework}
In the  3$\nu$ mixing scheme with massive neutrinos, $\chi_j$, being Majorana particles, the \betabeta-decay can be generated only by
the ($V-A$) charged current weak interaction via the exchange of the
three Majorana neutrinos $\chi_j$ having masses $m_j \lesssim$ a few
eV. The amplitude of the decay is proportional to the so-called Majorana effective mass
\cite{Bilenky:2001rz,Hirsch:2006tt,Rodejohann:2011mu}:
\begin{widetext}
\begin{equation}
\meff = \left| m_1\, |U_{\mathrm{e} 1}|^2
+ m_2\, |U_{\mathrm{e} 2}|^2\, e^{i\lambda_{21}}   
 + m_3 |U_{\mathrm{e} 3}|^2e^{i\lambda_{31}} \right|\,.
\label{effmass2}
\end{equation}
\end{widetext}

In eq. (\ref{effmass2}), $U_{ej}$, $j=1,2,3$  are the elements of the 
first row of the Pontecorvo-Maki-Nakagawa-Sakata (PMNS) matrix, $U$, describing 
the neutrino mixing \cite{pdg}, and
$\lambda_{21}$ and $\lambda_{31}$ are
the two Majorana CP violation (CPV) phases.\\
In the standard parametrization of the neutrino mixing matrix, we have that the elements, $U_{ej}$, depend only on
the reactor and the solar  neutrino mixing angles, $\theta_{13}$ and $\theta_{12}$
. More precisely these elements can be written as:
$$\begin{array}{l}|U_{\mathrm{e}1}| =\cos\theta_{12}\cos\theta_{13},\\
|U_{\mathrm{e}2}| =\sin\theta_{12}\cos\theta_{13}, \\
|U_{\mathrm{e}3}| =\sin\theta_{13}.\end{array}$$
Neutrino oscillation data are still compatible with two type of neutrino mass spectra:
i) normal ordered, $m_1<m_2<m_3$  and ii) inverted ordered,
$m_3<m_1<m_2$. The ordering of the masses determines a peculiar dependence of \meff with respect to  the lightest neutrino mass  and
therefore also with respect to the sum of the light active neutrinos $\sum_i m_i$.
Depending on the value of the lightest neutrino mass, $min(m_j)$, and on the hierarchy of the neutrino masses, the value of \meff can be:
\begin{itemize}
 \item \textit{quasi-degenerate (QD)}: \\
 $m_1\approxeq m_2 \approxeq m_3 \approxeq m_0 \gtrsim 0.1 eV$ gives $\meff \gtrsim0.05 eV$.
 \item \textit{inverted hierarchical (IH)}: \\
 $m_3 \ll m_1<m_2$ so $m_{1,2}\sim\sqrt{\Delta  m^2_{A}}$ gives $0.015\lesssim\meff\lesssim0.05 eV$.
  \item \textit{normal hierarchical (NH)}: \\
 $m_1 \ll m_2<m_3$ so
 $m_2\sim\sqrt{\Delta  m^2_{\odot}}$, $m_3\sim\sqrt{\Delta  m^2_{A}}$ and so $\meff\lesssim0.015eV$.
\end{itemize}
Further, the minimum for the sum of the neutrino masses compatible with current neutrino
oscillation data, namely when the lightest neutrino is massless, is $(\sumass)_{min}=5.87 \times10^{-2}$ eV for 
NH  and $(\sumass)_{min}=9.78 \times10^{-2}$ eV
for IH.

The \betabeta-decay experimental search is therefore compelling due to the
enormous impact in determining the nature of massive neutrinos, in
constraining the absolute neutrino mass scale as well as in the possibility to test the type of hierarchy for the neutrino masses.
So far, current experiments did not observe the decay. From those searches, one can derive
lower bounds on the half-lives of the utilized isotopes (see e.g. \cite{Piquemal:2013uaa} for a summary).
Combining the results by GERDA-I, the Hidelberg-Moscow and IGEX data sets, a limit on the 
$^{76}$Ge half-life was obtained $T_{1/2}^{0\nu}(^{76}$Ge$)>3.0\times 10^{25}$ yr at $90\%$ C.L.
\cite{Agostini:2013mba}. Limits for $T_{1/2}^{0\nu}(^{136}$Xe$)$ have been set by the EXO-200 and 
KamLAND-ZEN Collaborations \cite{Gando:2012zm,Albert:2014awa}.

The next generation of \betabeta-decay experiments ($\gtrsim 5$yr), such as  GERDA-II ($^{76}$Ge), SuperNEMO ($^{82}$Se)
CUORE ($^{130}$Te) and nEXO-200 ($^{136}$Xe) to name a few,
have been planned to reach a sensitivity of $T_{1/2}^{0\nu}\sim (1 - 1.5)\times 10^{26}$ yr. 
This is meant to explore the inverted hierarchical region for $\meff$, see the dashed region in
Fig. (\ref{fig:meffpseudoIH}). Since  $\meff$ scales like $\sqrt{T_{1/2}^{0\nu}}$
one can expect to push down $\meff$ of a factor of 3 or 4 with respect  to the present lower bounds.
Of course this could be not enough to cover the entire IH region, so at the moment many efforts are 
focused to reach  values for \meff around $10$ meV.
In this case detector technology plays a crucial role (a reasonable signal-to-background ratio
in the energy interval $\Delta  E$ of approximately a full width at half maximum (FWHM) around
the Q-value is indeed related to possible high
resolution detectors) together with
the choice of the  isotope used in the experiment.

Even though  for the current  running and near-term  \betabeta-decay experiments will
probably be difficult to  explore the entire IH region  it makes
sense, from a theoretical point of view, to consider potential
\betabeta-decay experiments aiming to measure half-lives of the order of $10^{27}$ yrs or more.
This generation of experiments,  which we will label in the following as 
\emph{mega}-experiments for $T_{1/2}^{0\nu}\sim10^{27}$  yrs and \emph{ultimate}-experiments for $T_{1/2}^{0\nu}\sim10^{29}$  yrs \cite{Dell'Oro:2014yca}, could test
several scenarios involving Majorana neutrinos.

The \betabeta-decay experimental searches are also complementary  to $\beta$-decay and cosmology, 
which has now entered its precision era \cite{Minakata:2014jba,Dodelson:2014tga}.  The  constraints given by the Planck Collaboration
on the sum of the light active neutrinos which were  obtained from the measurements of the 
cosmic microwave background (CMB) temperature and lensing-potential power spectra 
make tantalizing this kind of combined analysis. We will consider as reference value the limit \cite{Ade:2013zuv}

\begin{equation}
\sum_i m_i \leq 0.23 \mbox{eV\, 95\% C.L.}
\label{eq:Planck}
\end{equation}

Future large scale structure surveys like the recently approved EUCLID \cite{Laureijs:2011mu}, 
will allow to constrain  $\sum_i m_i$ up to few meV if is combined with the Planck data.
We will consider in the following  a future sensitivity of $\sumass\lesssim$ 0.15 eV. 

The complementarity of $\meff$ and the constraints from cosmological observables
is starting to be relevant in scenarios or models for neutrino physics\footnote{See for 
instance the case of zero textures \cite{Meloni:2014yea} and neutrino mass sum rules 
\cite{Barry:2010yk,Dorame:2011eb,King:2013psa}.}. Moreover, the limits from cosmology for the neutrino masses will be complemented by the the Tritium-beta decay experiment KATRIN~\cite{Bornschein:2003xi,fortheKATRIN:2013saa} whose sensitivity for the neutrino mass is expected to be around $0.2~eV$.
 
In this paper we explore testable scenarios in view of the next  {\it mega-} and {\it ultimate}-generation of 
$\betabeta$-decay searches such as the admixture of pseudo-Dirac and Majorana neutrinos. 
Along the line with previous works (see e.g. \cite{Barry:2010en})
we explore how the standard predictions for \betabeta-decay are  considerably modified 
with the most recent neutrino oscillation data
 \cite{Forero:2014bxa,GonzalezGarcia:2012sz,Capozzi:2013csa}
and especially
we investigate which of these scenarios  can be 
experimentally reachable in the {\it mega-} and {\it ultimate}-generations even for the NH spectrum (in the standard case in fact
strong cancellations could occur leading to a zero \meff). 
 Further, it is also known that in the presence 
of extra light sterile neutrinos, with square mass difference of the order of $1 ~eV^2$, the 
$\meff$ is also modified \cite{Barry:2011wb, Girardi:2013zra}\footnote{For the infuence of a keV sterile neutrino on the neutrino-less
double beta decay effective mass see~\cite{Merle:2013ibc}}. We will show how the analysis for 
\betabeta-decay and pseudo-Dirac neutrinos
changes in the presence of one extra sterile state.
\\
\section{Light Pseudo-Dirac Neutrinos}

         As it is well known in the three Majorana neutrino framework  the \betabeta-decay amplitude
          can well be zero in the NH case due to internal cancellations operated by  the two Majorana phases
          spanning values in the range $[0,2\pi]$. This behavior is not present in the IH case \cite{Bilenky:2001rz}.
          Further, in the NH scheme, uncertainties in the parameters
          governing neutrino oscillations, namely the angles and the two square mass differences,
          determine  the existence of a quite large interval for the absolute neutrino mass scale
          in which \meff can be zero.
          In the case in which at least one of the light active neutrinos $\chi_i$, $i=1,2,3$ 
          is a Dirac fermion, such field
          (which is the field with definite mass eigenstate) 
          does not participate in the amplitude of the \betabeta-decay 
          since it is distinct from its antiparticle
          and thus it is not possible to Wick-contract the neutrino field operator in order to
          obtain a neutrino propagator \cite{Petcov:1982ya,valle:1983yw,Leung:1983ti}.

          
          In this framework neutrinos are usually referred as pseudo-Dirac particles
         ---when the two neutrinos are active-sterile we have the so-called 
          quasi-Dirac neutrino \cite{valle:1983yw} and when they are active-active we have the so called
          pseudo-Dirac neutrino \cite{Wolfenstein:1981kw}\footnote{For examples of models see for 
          instance \cite{Barry:2010en,Machado:2010ui,Morisi:2011ge}.}.
          In this scenario the prediction for the \betabeta-decay half-life and thus for \meff can
          deeply be modified.

\begin{center}
\begin{figure}[h!]
 \subfigure
{\includegraphics[width=5cm,bb= 128 395 476 705]{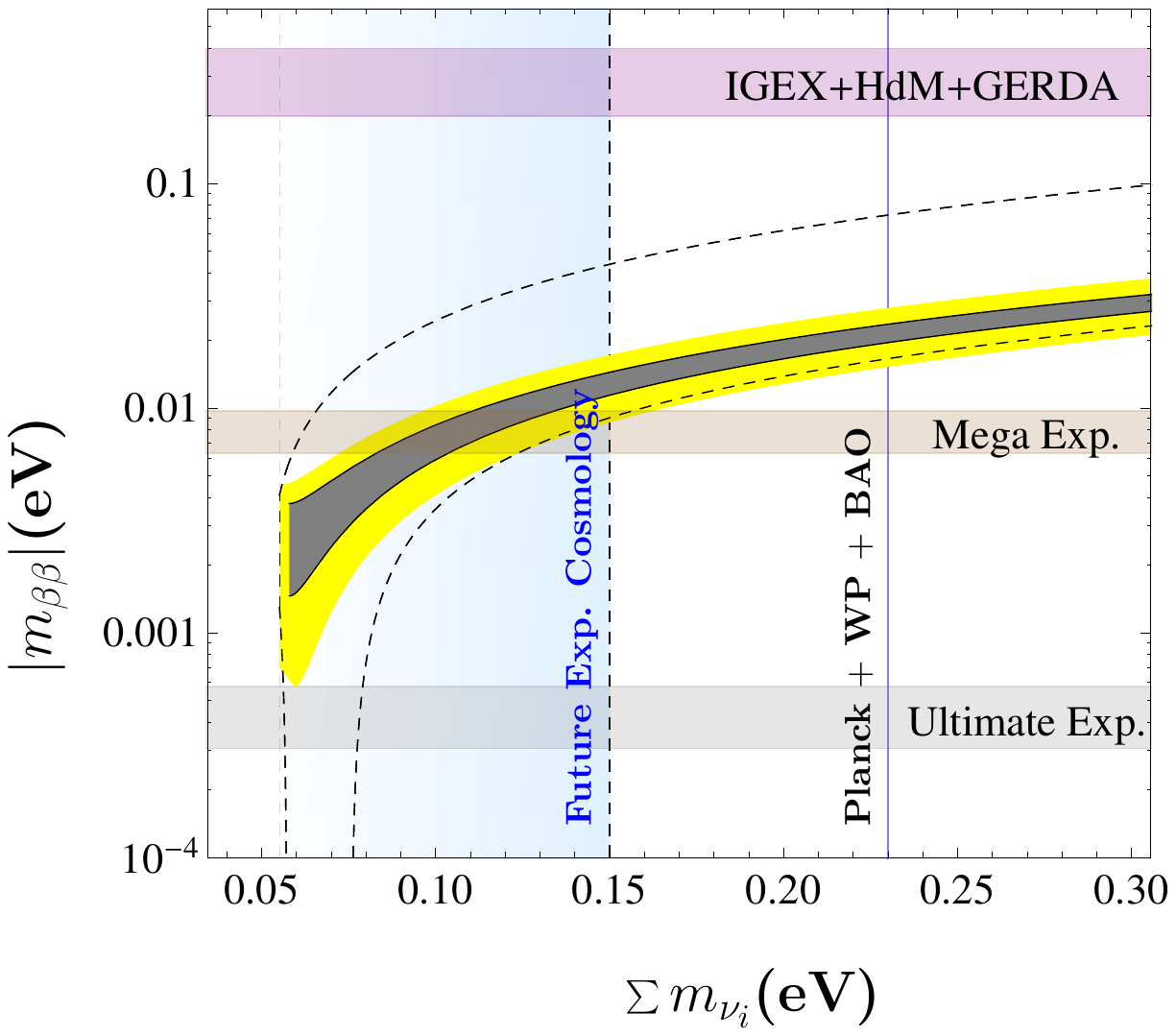}}
 \subfigure
 {\includegraphics[width=5cm,bb= 128 395 476 705]{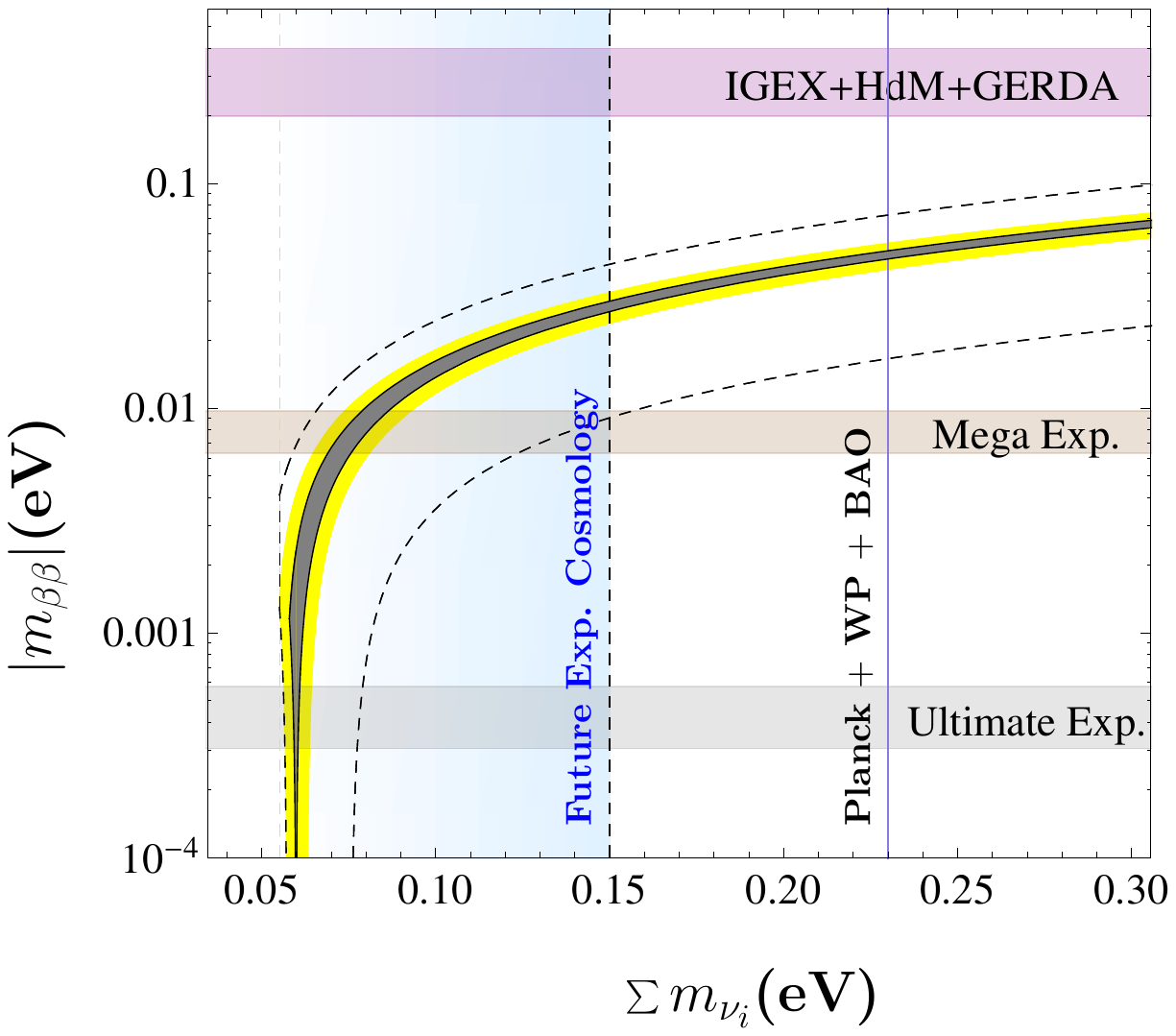}}
 \subfigure
 {\includegraphics[width=5cm,bb= 128 395 476 705]{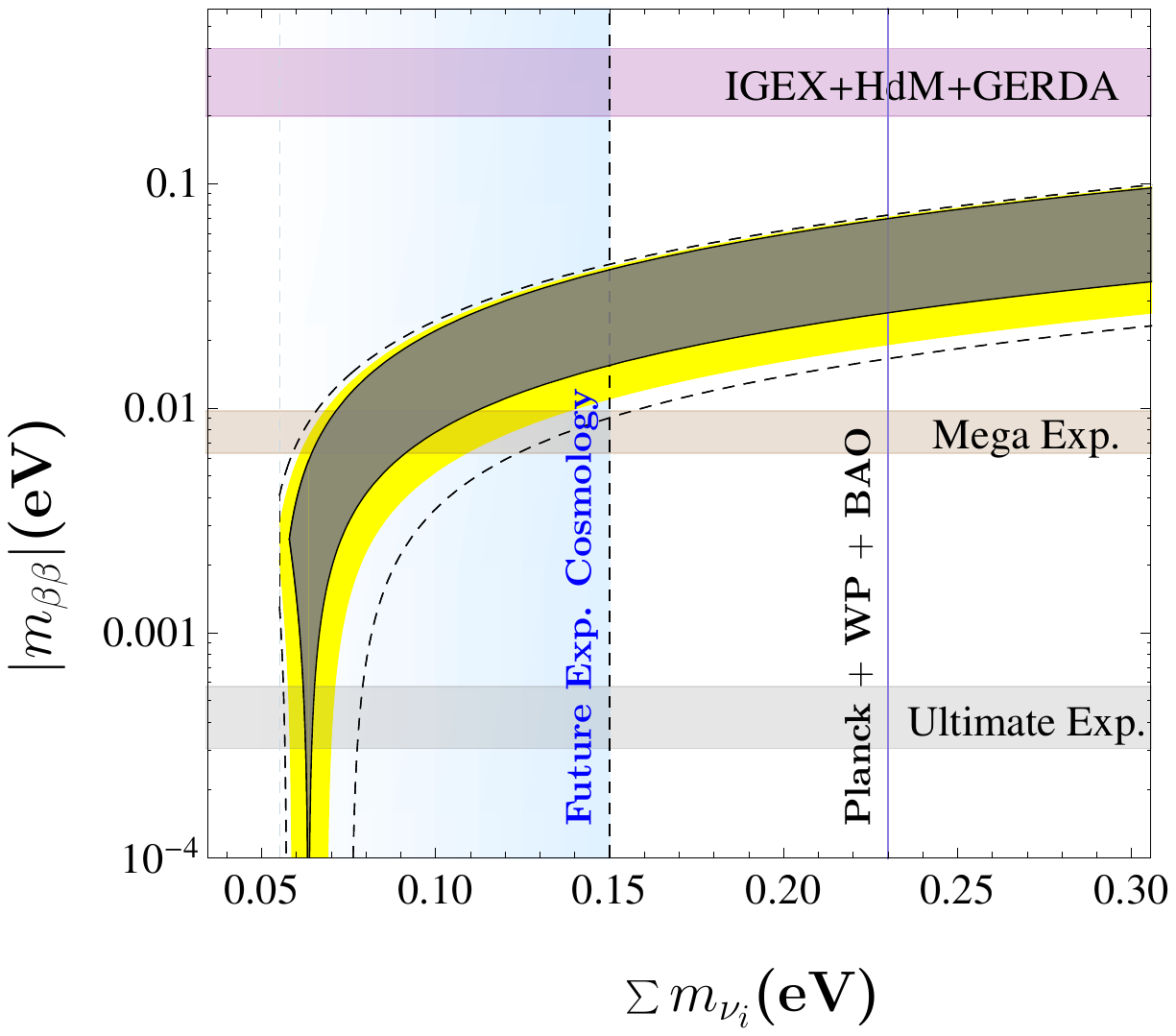}}
     \vspace*{-10pt}\caption{\label{fig:meffpseudo}Plots for \meff vs $\sumass$ for the NH case
    in the case in which respectively, from
    the left to the right,    $\chi_1$ or $\chi_2$ or $\chi_3$ is a Dirac fermion and therefore
    the related contribution is not present in \betabeta-decay amplitude. 
    The light (Yellow) region corresponds to a 3$\sigma$ uncertainty in the oscillation parameters
    while the  dark (Gray) area indicates the prediction for \meff for the best fit values  given in \cite{Capozzi:2013csa}.
    The solid vertical line
    corresponds to the Planck constraint in eq. (\ref{eq:Planck}). The shaded area for $\sumass\lesssim$ 0.15 eV
    indicates the future cosmological sensitivity. The dashed contour indicate the standard  predictions 
    with three Majorana active neutrinos.}
\end{figure}
\end{center}

\begin{center}
\begin{figure}[h!]
 \subfigure
 {\includegraphics[width=5cm,bb= 128 395 476 705]{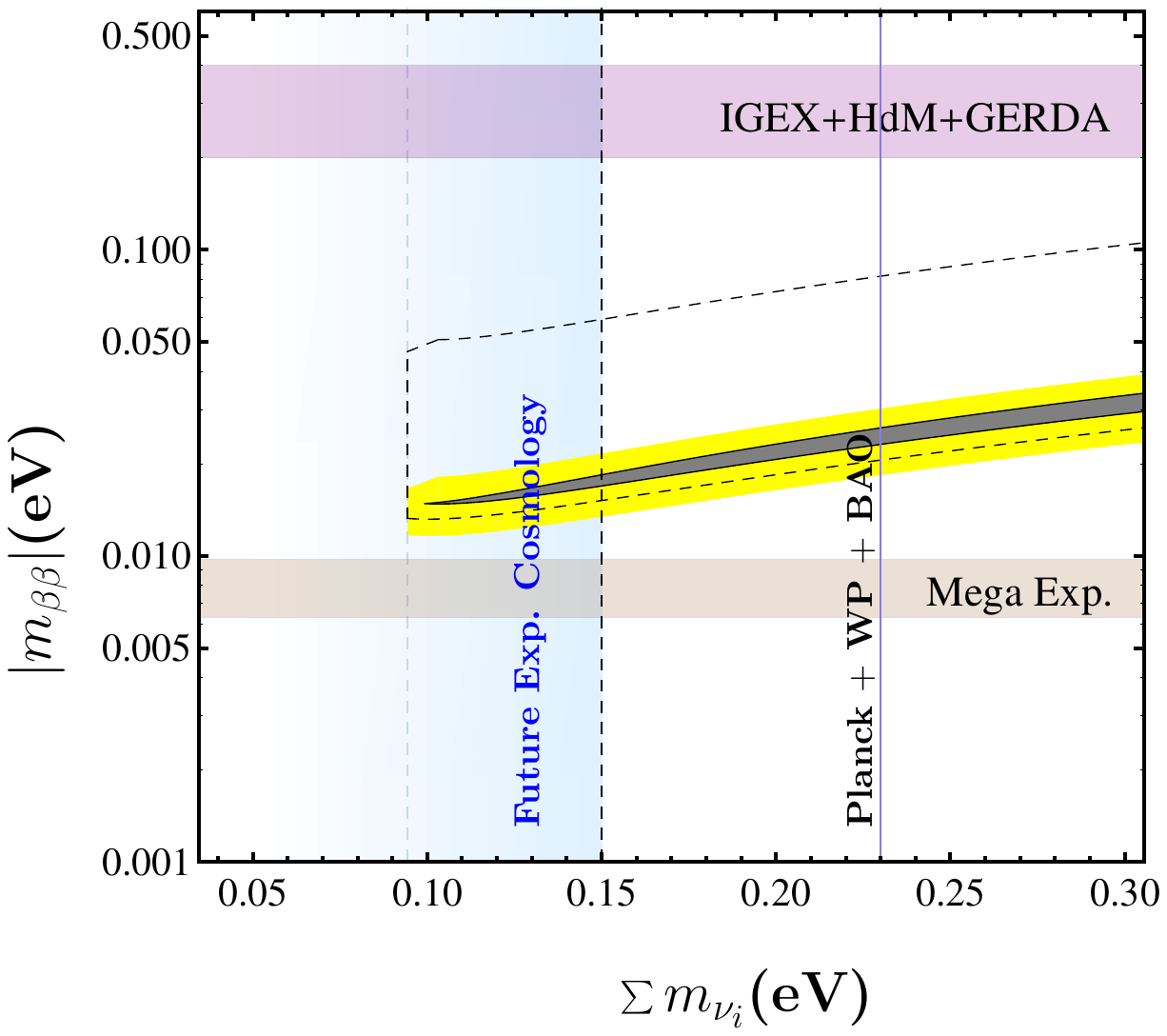}}
 \subfigure
 {\includegraphics[width=5cm,bb= 128 395 476 705]{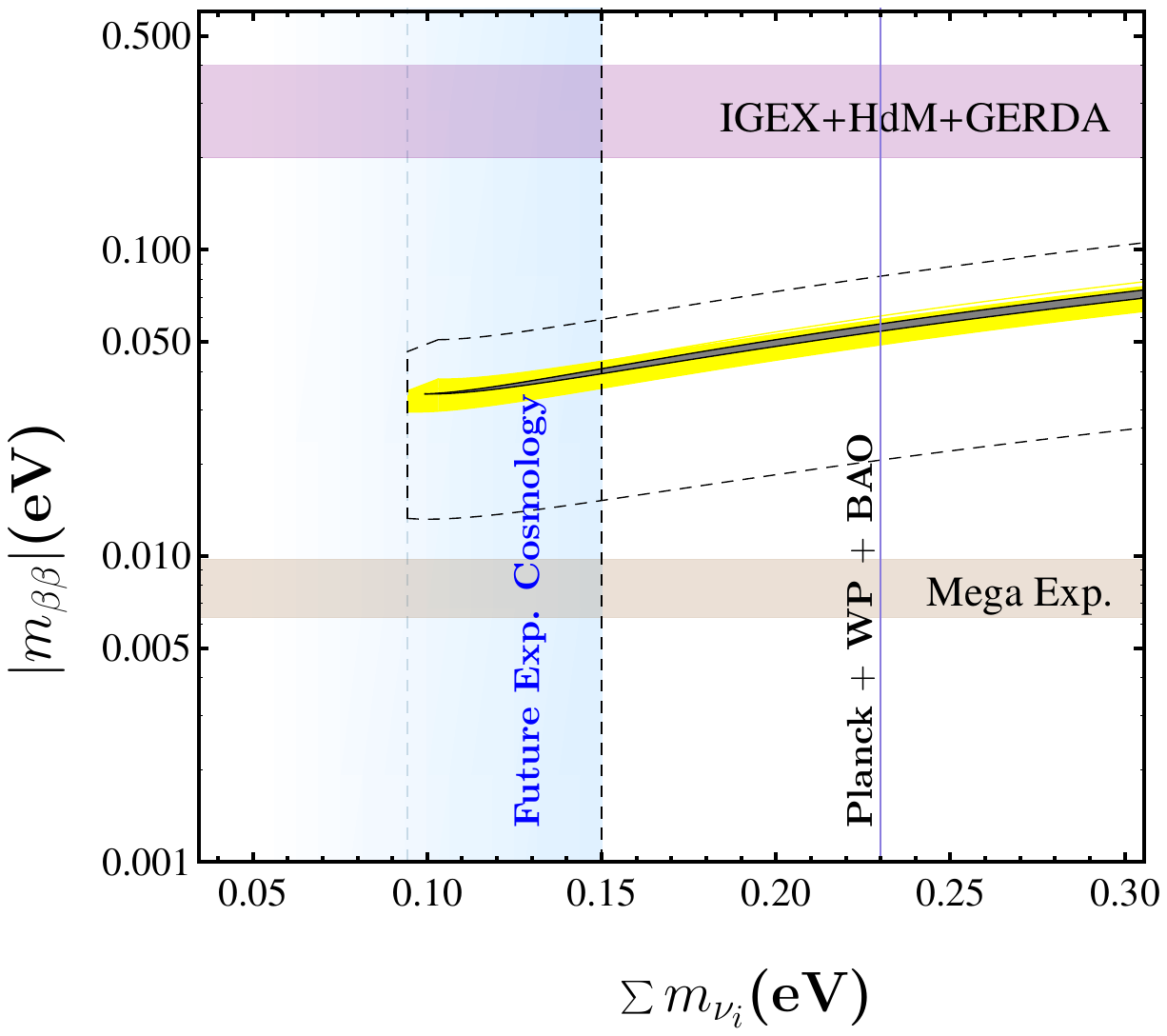}}
 \subfigure
 {\includegraphics[width=5cm,bb= 128 395 476 705]{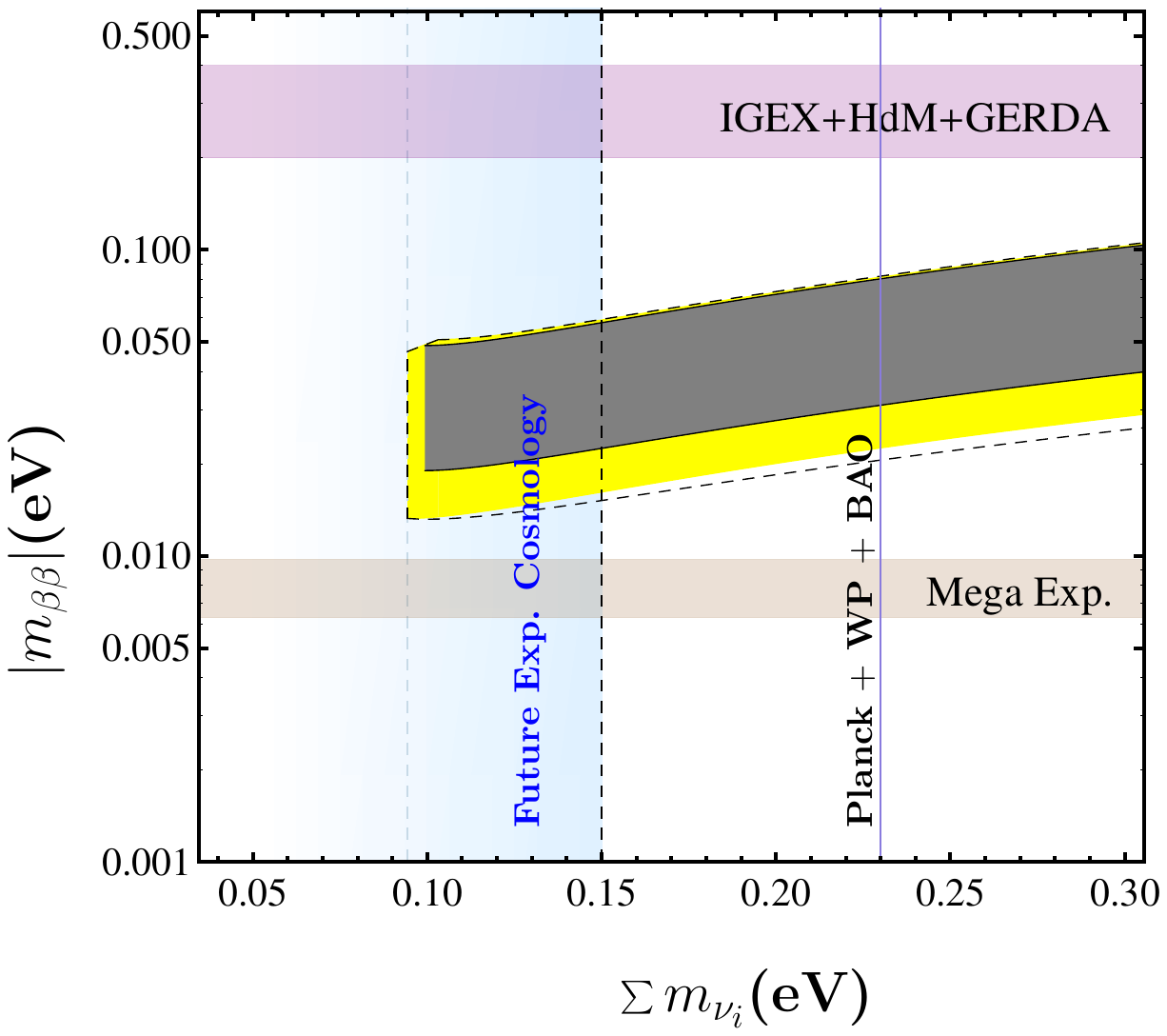}}
     \vspace*{-10pt}\caption{\label{fig:meffpseudoIH} Plots for \meff vs $\sumass$ for the IH case
    in the case in which respectively, from
    the left to the right,    $\chi_1$ or $\chi_2$ or $\chi_3$ is a Dirac fermion.
    The dashed contour indicates the standard 
    predictions with three Majorana active neutrinos. The same colors of Fig. (\ref{fig:meffpseudo}) 
    are used for the allowed regions.}
\end{figure}
\end{center}

         In Fig. (\ref{fig:meffpseudo}) and (\ref{fig:meffpseudoIH}) we show 
         the dependence of \meff with respect to the sum of the light active states $\sumass$ in the case of NH and IH
         respectively if only one of the neutrinos $\chi_i$ is a Dirac fermion. 
         In both plots we show the dashed contour which corresponds to the general allowed 3$\sigma$ region
         in the standard 3 Majorana neutrino case. The light (Yellow) region corresponds to a
         3$\sigma$ uncertainty in the oscillation parameters  while the 
         dark (Gray) area indicates the prediction for \meff for the best fit values in ref. \cite{Capozzi:2013csa}. 
         In the NH case it is clear from the plots that 
         if $\chi_1$ is a Dirac particle then a lower limit for \meff is found
         (differently from \cite{Barry:2010en} the current value for $\theta_{13}$ determines a non vanishing \meff). 
         If instead $\chi_2$ (or $\chi_3$)
         are Dirac then a cancellation is still possible especially for 
         $\sumass$ quite near the lower limit given by the current oscillation 
         experiments. 
         The vertical solid line on those plots represents the Planck+WP+BAO limit given in eq. (\ref{eq:Planck})
         while the vertical dashed line delimiting the shaded area is 
         indicating the sensitivity of the next-generation of cosmological surveys i.e. $\sumass\lesssim 0.15$ eV.
         The three horizontal bands indicate the GERDA-I limit and the two ranges 
         for \meff corresponding to the {\it mega-} and {\it ultimate}-\betabeta-decay experiments 
         (the intervals for \meff have been computed using the NMEs used in \cite{Meroni:2012qf}).
         For the IH scenario in Fig. (\ref{fig:meffpseudoIH})
         there is no complete cancellation for the \meff (as in the usual
         three Majorana neutrino case),  so  all these scenarios can 
         be falsified in the next generation of \betabeta-decay experiments.

\begin{center}
\begin{figure}[h!]
 \subfigure
 {\includegraphics[width=5cm,bb= 128 395 476 705]{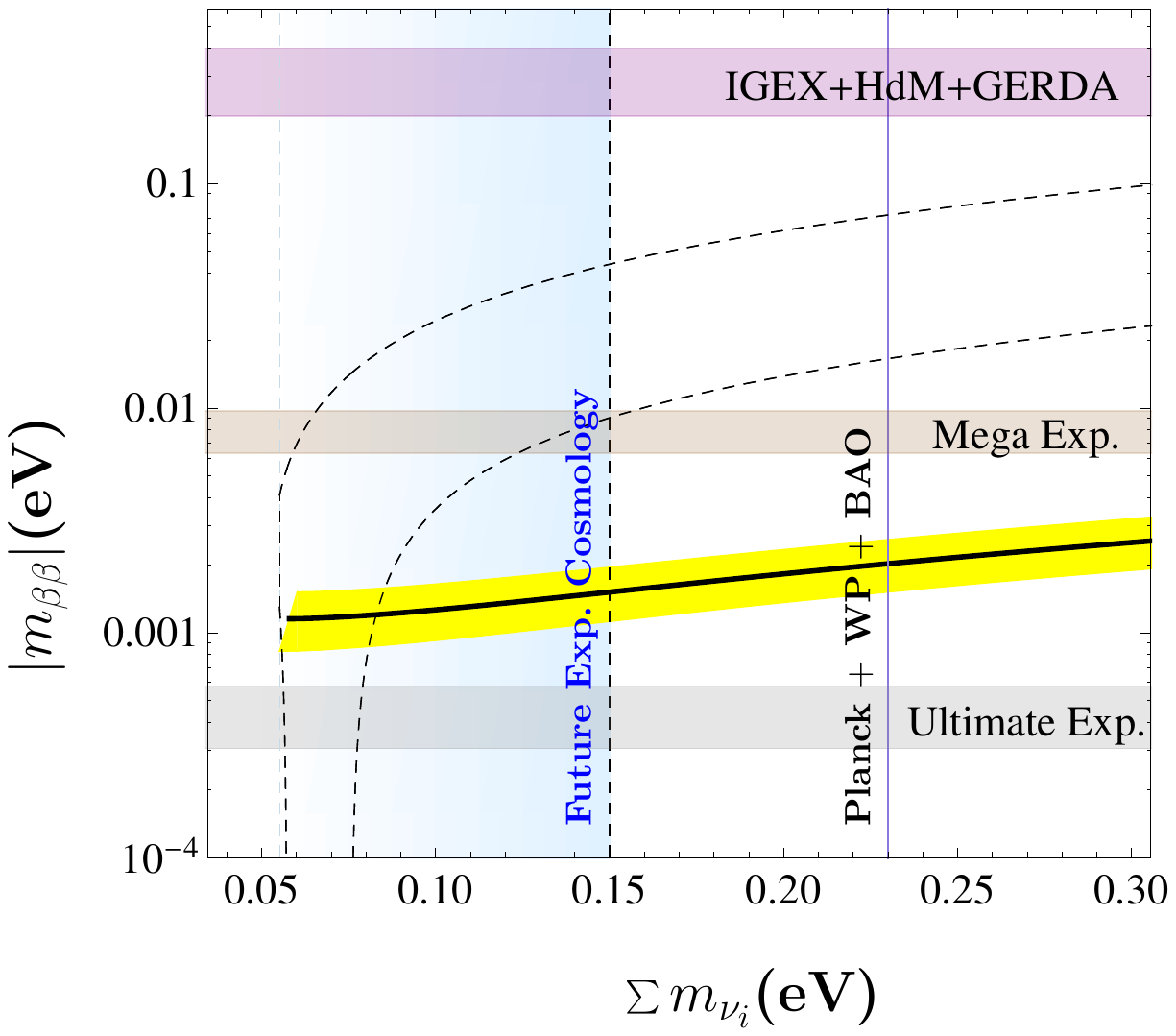}}
 \subfigure
 {\includegraphics[width=5cm,bb= 128 395 476 705]{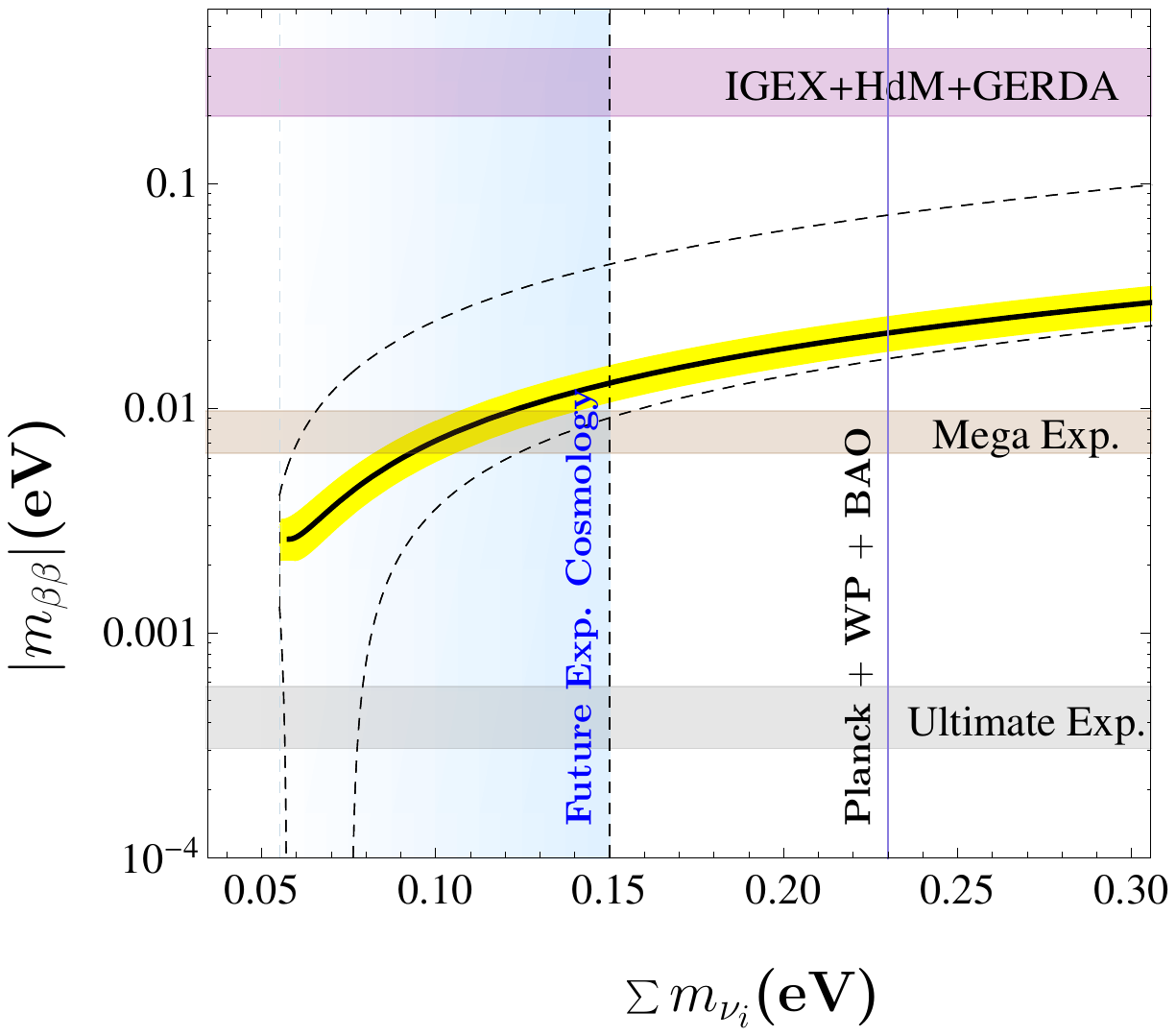}}
  \subfigure
 {\includegraphics[width=5cm,bb= 128 395 476 705]{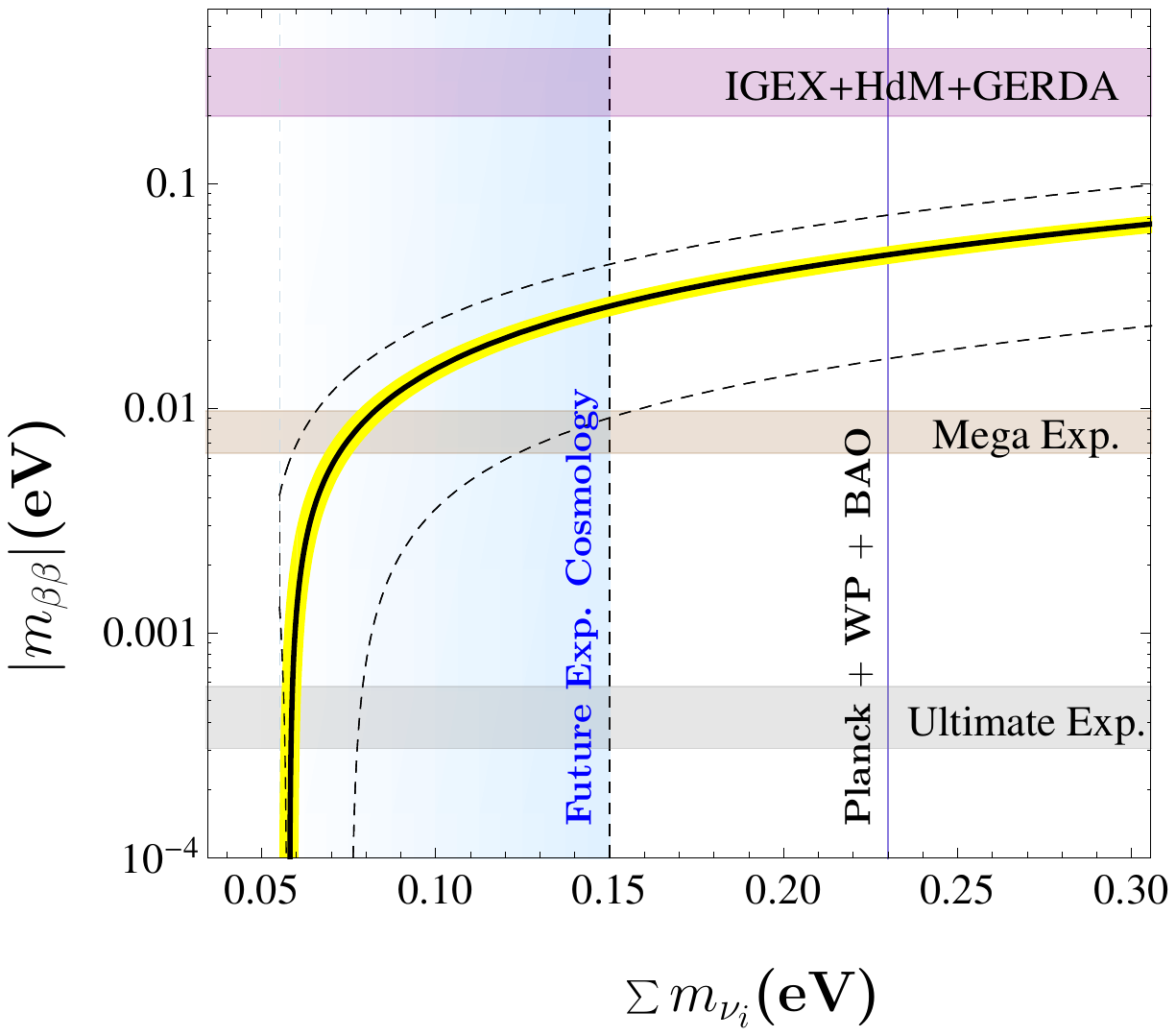}}
      \vspace*{-10pt}\caption{\label{fig:meffpseudo2}Plots for \meff vs $\sumass$ for the NH case
    in the case in which respectively, from
    the left to the right,    $(\chi_1,\chi_2)$ or $(\chi_1,\chi_3)$ 
    or $(\chi_2,\chi_3)$ are Dirac fermions. The same colors of Fig. (\ref{fig:meffpseudo}) 
    are used for the allowed regions.}
\end{figure}
\end{center}

\begin{center}
\begin{figure}[h!]
 \subfigure
 {\includegraphics[width=5cm,bb= 128 395 476 705]{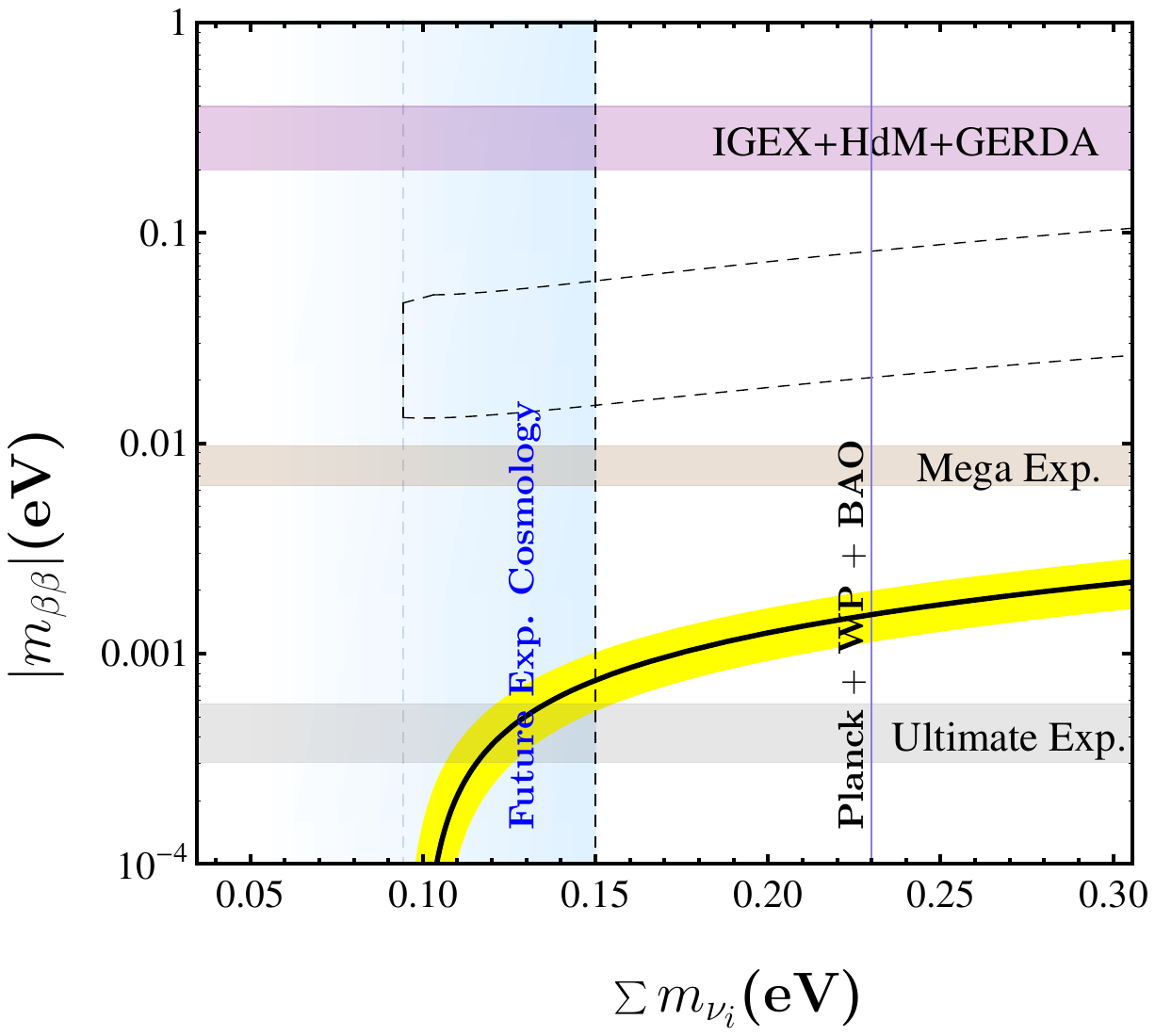}}
 \subfigure
 {\includegraphics[width=5cm,bb= 128 395 476 705]{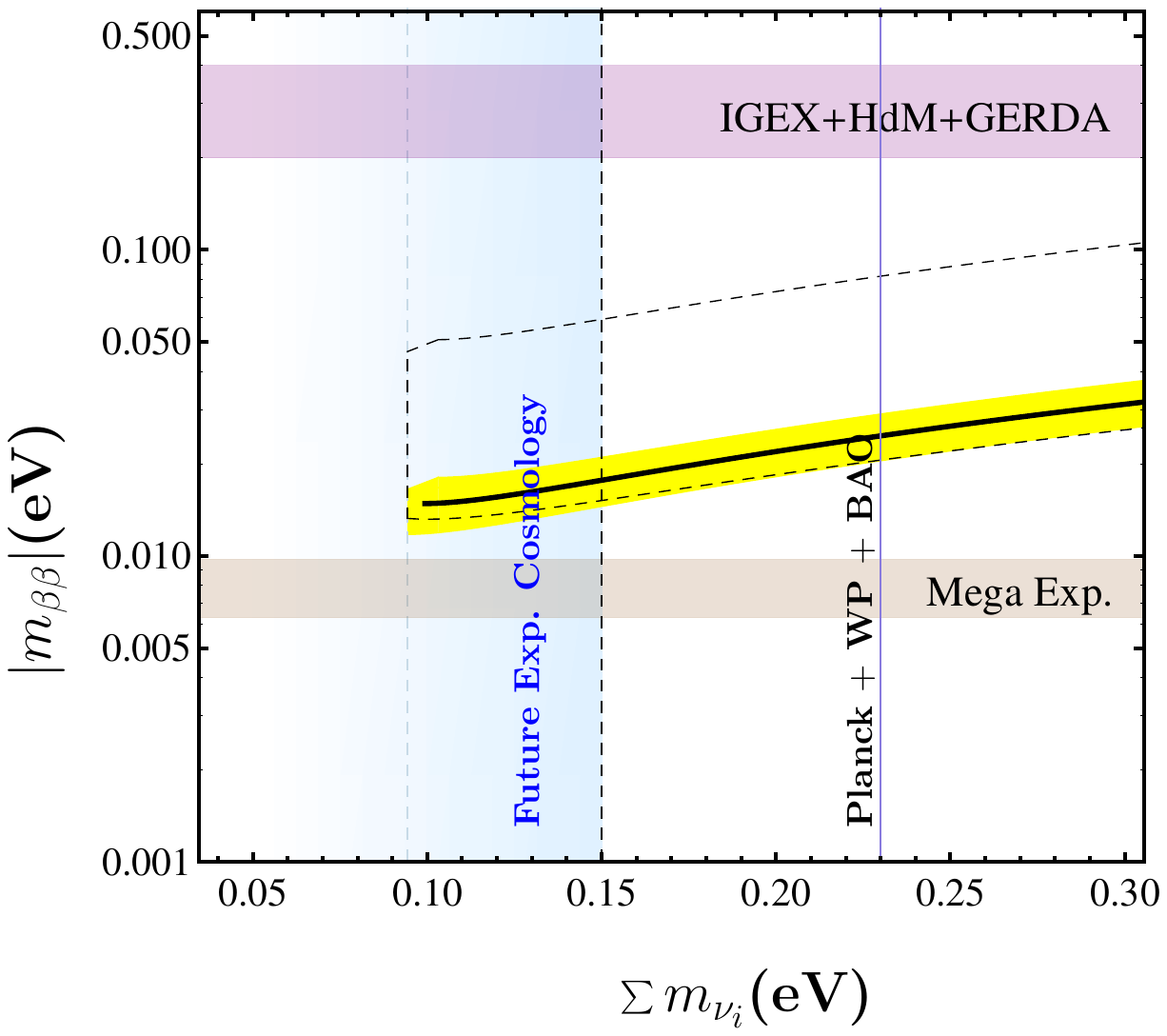}}
  \subfigure
 {\includegraphics[width=5cm,bb= 128 395 476 705]{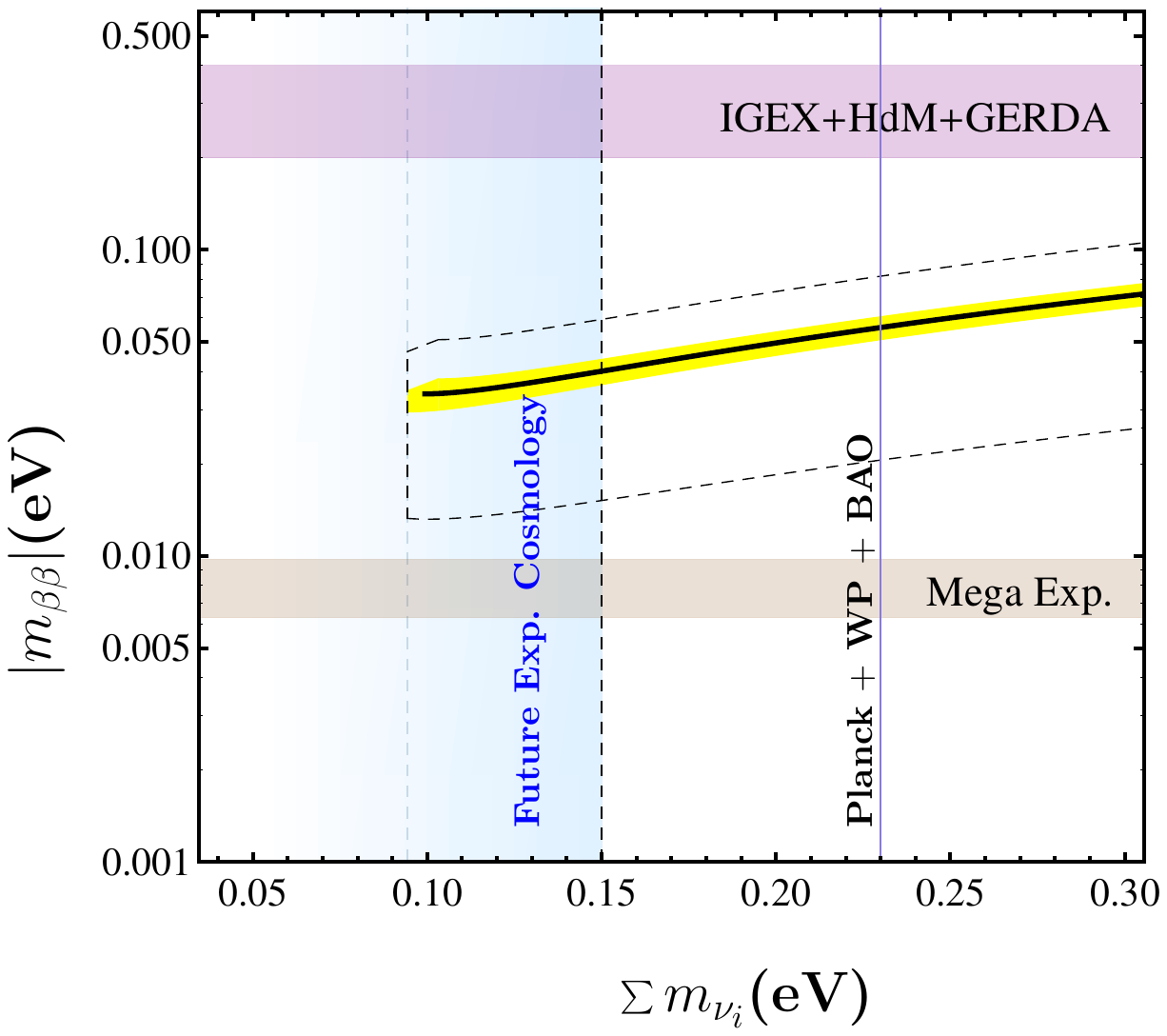}}
      \vspace*{-10pt}\caption{\label{fig:meffpseudo2IH}Plots for \meff vs $\sumass$ for the IH case
    in the case in which respectively, from
    the left to the right,    $(\chi_1,\chi_2)$ or $(\chi_1,\chi_3)$ 
    or $(\chi_2,\chi_3)$ are Dirac fermions. The same colors of Fig. (\ref{fig:meffpseudo}) 
    are used for the allowed regions.}
\end{figure}
\end{center}

        The case when two of the active states are Dirac particles is represented in the plots in Fig. (\ref{fig:meffpseudo2}) for NH and in Fig. (\ref{fig:meffpseudo2IH}) for IH.
        Here again the light (Yellow) region corresponds to a 3$\sigma$ uncertainty in the oscillation parameters  while the 
	thick solid line is determined with the best fit values in ref. \cite{Capozzi:2013csa}. The NH cases can all be tested in the \textit{mega}-generation 
(in the case $\chi_2$ and $\chi_3$ are both Dirac then the interval of $\meff\sim 0$ corresponds to
the limit of a massless neutrino). In the case of IH if $(\chi_1,\chi_3)$ 
    or $(\chi_2,\chi_3)$ are Dirac fermions then they can be also tested by the \textit{mega}-generation 
    while the case of  $(\chi_1,\chi_2)$ will be very difficult if not impossible to rule out.
\\
\section{Extra Sterile Neutrinos}\label{sterile}

In the past decade the framework of 3-neutrino mixing has been challenged by 
the existence of experimental  anomalies found in the 
i) results of the LSND \cite{Aguilar:2001ty} and MiniBooNE experiments \cite{Aguilar-Arevalo:2013pmq}
ii) re-analyses of the short baseline (SBL) reactor neutrino oscillation data using newly calculated fluxes of reactor
$\bar{\nu}_e$, which detect a ``disappearance''
of the reactor $\bar{\nu}_e$ (``reactor neutrino anomaly'') \cite{Huber:2011wv,Mueller:2011nm}
iii) results of the calibration experiments of
the radio-chemical Gallium solar neutrino detectors 
GALLEX and SAGE (``Gallium anomaly'')\cite{Kaether:2010ag, Abdurashitov:2009tn} ---for a review see e.g.
\cite{Abazajian:2012ys}.
These anomalies could be explained enlarging the  field content in the neutrino sector  
with  one or more massive neutrinos
with a square mass difference  around 1 eV$^2$.
  
  \vspace*{-20pt}
\onecolumngrid
\vspace*{-2cm}
\vspace*{-2cm}

\onecolumngrid

\vspace*{-50pt}
\twocolumngrid
It is worth stressing that  if this extra sterile state exists
it  would be completely thermalised in the early universe through 
mixing and scattering processes. However this is  not confirmed by recent cosmological results
which actually seem to disfavour these scenarios
since such sterile state(s) would lead to a too strong suppression of
structure formation.
Nevertheless sterile states with $1$ eV mass,
as indicated by SBL data, are still possible if either some neutrino-antineutrino asymmetry
is at work in order to  inhibit the
sterile neutrino production in the early universe 
or a modification of the cosmological model is provided.
In any case a mechanism  preventing a full thermalization is needed (see
\cite{Archidiacono:2014apa} and references therein).
It has been noticed in some works (see e.g. \cite{Barry:2011wb, Girardi:2013zra}) that if one or two sterile
Majorana neutrinos  exist then they deeply modify the predictions for
the \betabeta-decay.
In the simplest scheme with one extra sterile neutrino, the so called $3+1$ scheme,
the matrix which describes neutrino mixing is now a  $4\times4$ unitary matrix
which can be written as:
\begin{equation}\begin{split}
U &=  O_{34} V_{24} O_{23} O_{14} V_{13} V_{12}\,\, \\
&\times diag (1,e^{i\lambda_1},e^{i\lambda_2},e^{i\lambda_3}),
\end{split}
\end{equation}
where $O_{ij}$ and $V_{kl}$ are real and complex rotations in
$i-j$ and $k-l$ planes respectively, while $\lambda_1$, $\lambda_2$ and
$\lambda_3$ are three CPV Majorana phases. In the effective Majorana mass \meff only the elements of the first row of the 
neutrino mixing matrix are relevant and their expressions can be written in a completely generic way as:
\begin{equation}\begin{array}{ll}
 U_{e1}  = c_{12} c_{13} c_{14},  &\quad U_{e2}  = e^{i \alpha/2} c_{13} c_{14} s_{12}, \\ \\
U_{e3}  = e^{i\beta/2} c_{14} s_{13}, &\quad U_{e4}  = e^{i\gamma/2} s_{14}\,,
\end{array}\label{U31}
\end{equation}
%
where we have used the standard notation
$c_{ij} \equiv \cos\theta_{ij}$ and
$s_{ij} \equiv \sin\theta_{ij}$ and $\alpha,\beta$ and $\gamma$ are a definite combination
of the phases appearing in $U$.
The element $U_{e4}$, and thus the angle
$\theta_{14}$, describes the coupling of fourth
neutrino $\chi_4$ to the electron in the
weak charged lepton current.
The \meff
in this framework  has the form:

\begin{widetext}
\begin{equation}
\meff =\left|  m_1 |U_{e1}|^2 +  m_2 |U_{e2}|^2 e^{i \alpha} +
m_3 |U_{e3}|^2e^{i \beta} + m_4 |U_{e4}|^2 e^{i \gamma}
\right|.
\end{equation}

Depending on the neutrino mass hierarchy, we will have:
\begin{itemize}
 
\item  NH:
$$  \qquad
\begin{array}{lr}
m_1\equiv m, &m_2=\sqrt{m^2+\Delta m^2_{21}},\\ \\
 m_3=\sqrt{m^2+\Delta m^2_{21}+\Delta m^2_{31}},&m_4=\sqrt{m^2+ \Delta m^2_{41}}
\end{array}
$$ 
 \item IH: 
$$\begin{array}{lr}
m_1\equiv \sqrt{m^2+\Delta m^2_{31}},&m_2=\sqrt{m^2+\Delta m^2_{21}+\Delta m^2_{31}}, \\ \\
 m_3=m&m_4=\sqrt{m^2+ \Delta m^2_{41}}\end{array}
$$ 
\end{itemize}
\end{widetext}

\begin{figure}[h!]
 \subfigure
 {\includegraphics[width=5cm,bb= 128 395 476 705]{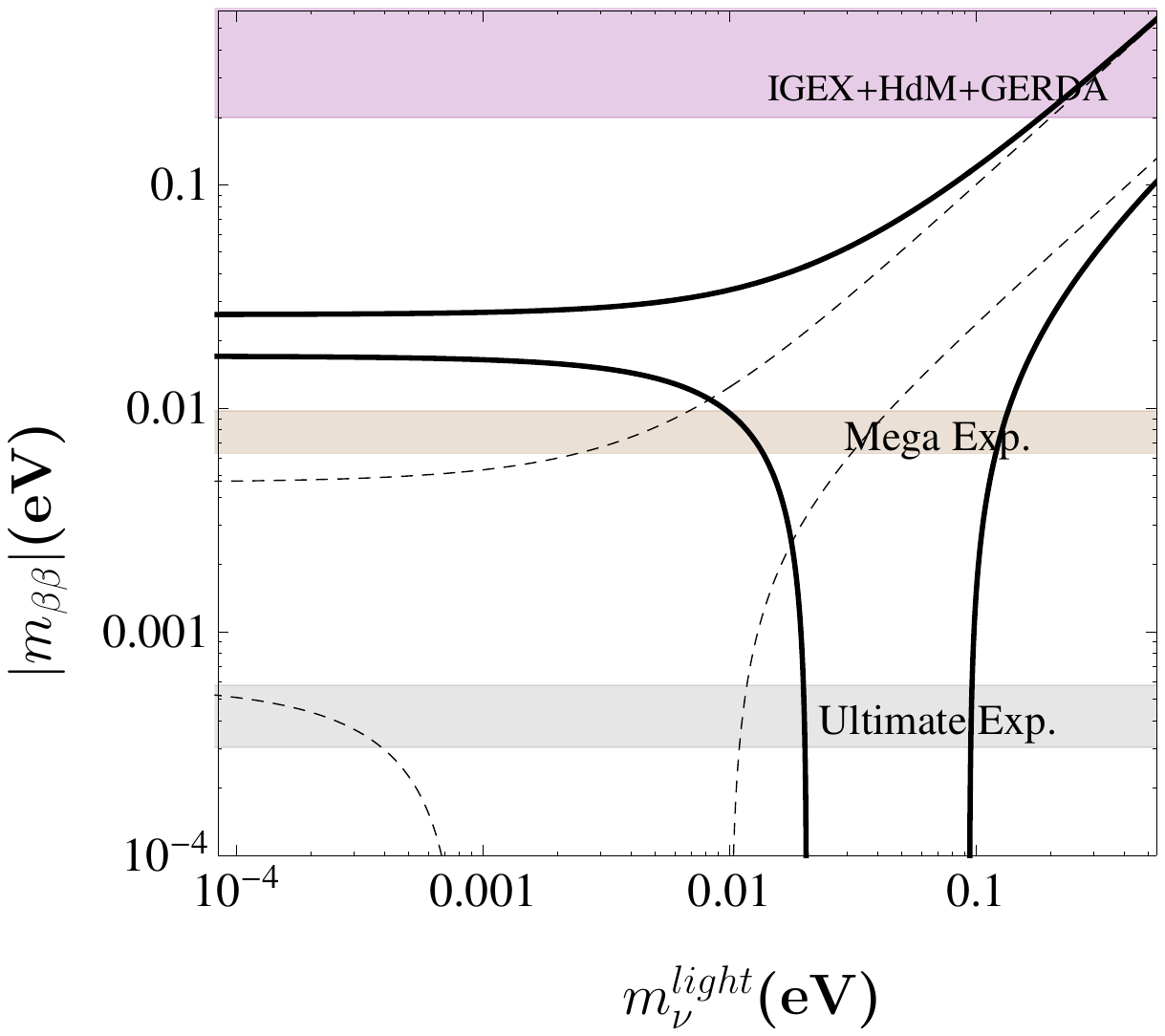}}
 \hspace{1cm}\subfigure
 {\includegraphics[width=5cm,bb= 128 395 476 705]{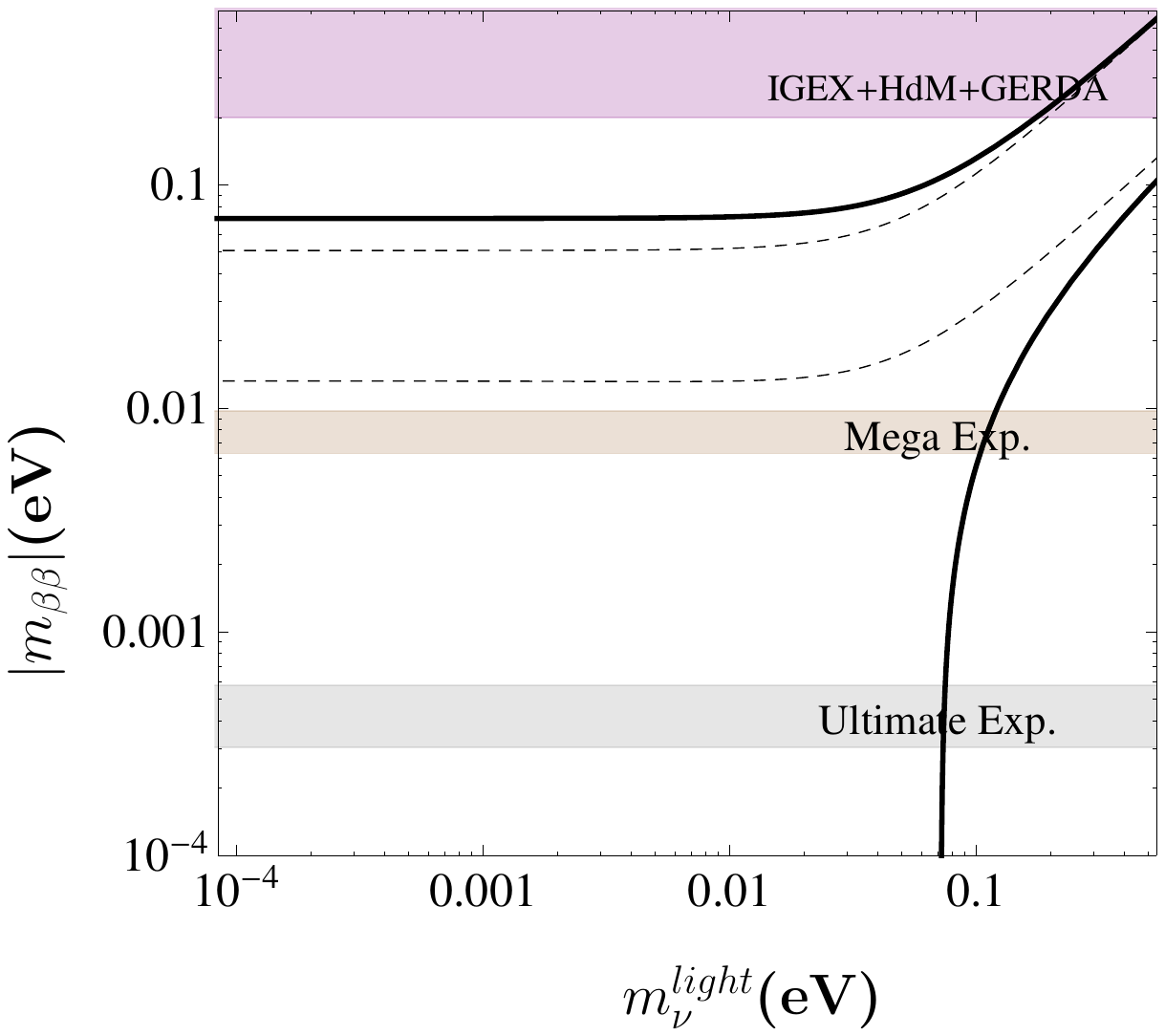}}
      \vspace*{-10pt}\caption{\label{fig:meffsterile}Plots for \meff vs $m_{\nu}^{light}$ 
    in the scenario with 1 extra sterile state with $\Delta m^2_{SBL}=0.93 eV^2$ and 
    $\sin\theta_{14}=0.15$ for NH (left panel) and IH (right panel). The dotted contours 
    define the allowed range for \meff in the 3$\nu$ 
    framework. See text for further details.}
    \end{figure}

In this section we want to illustrate the predictions for \meff 
in the scenario with one extra Majorana sterile state, $\chi_4$, and 
three light active states $\chi_1,\chi_2$, $\chi_3$ which can be pseudo-Dirac or Majorana particles.
We will consider as reference value for $\sin\theta_{14}$
and $\Delta m^2_{41}$ the result obtained in the global analyses performed in
\cite{Kopp:2013vaa}\footnote{For another global fit see \cite{Giunti:2013aea}. 
Here we only give an example of how the $\meff$ is modified in the scenario of 
pseudo-Dirac neutrinos and one Majorana sterile state.}:

\begin{equation}
\sin\theta_{14}= 0.15\,,~~~\Delta m^2_{41} = 0.93~{\rm eV^2}\,
\label{values}
\end{equation}
In Fig. (\ref{fig:meffsterile}) we show the \meff versus the lightest neutrino mass $m_\nu^{light}$ in the standard
3$\nu$ framework. In this case for both NH and IH the \meff can vanish. Moreover for small absolute masses,
namely for $m_\nu^{light}< 0.01$ eV and $\meff< 0.01$ eV only IH is possible therefore only the \textit{mega}-
and \textit{ultimate}-generation of \betabeta-decay experiments 
might constrain this scenario.

\begin{center}
\begin{figure}[h!]
 \subfigure
 {\includegraphics[width=5cm,bb= 128 395 476 705]{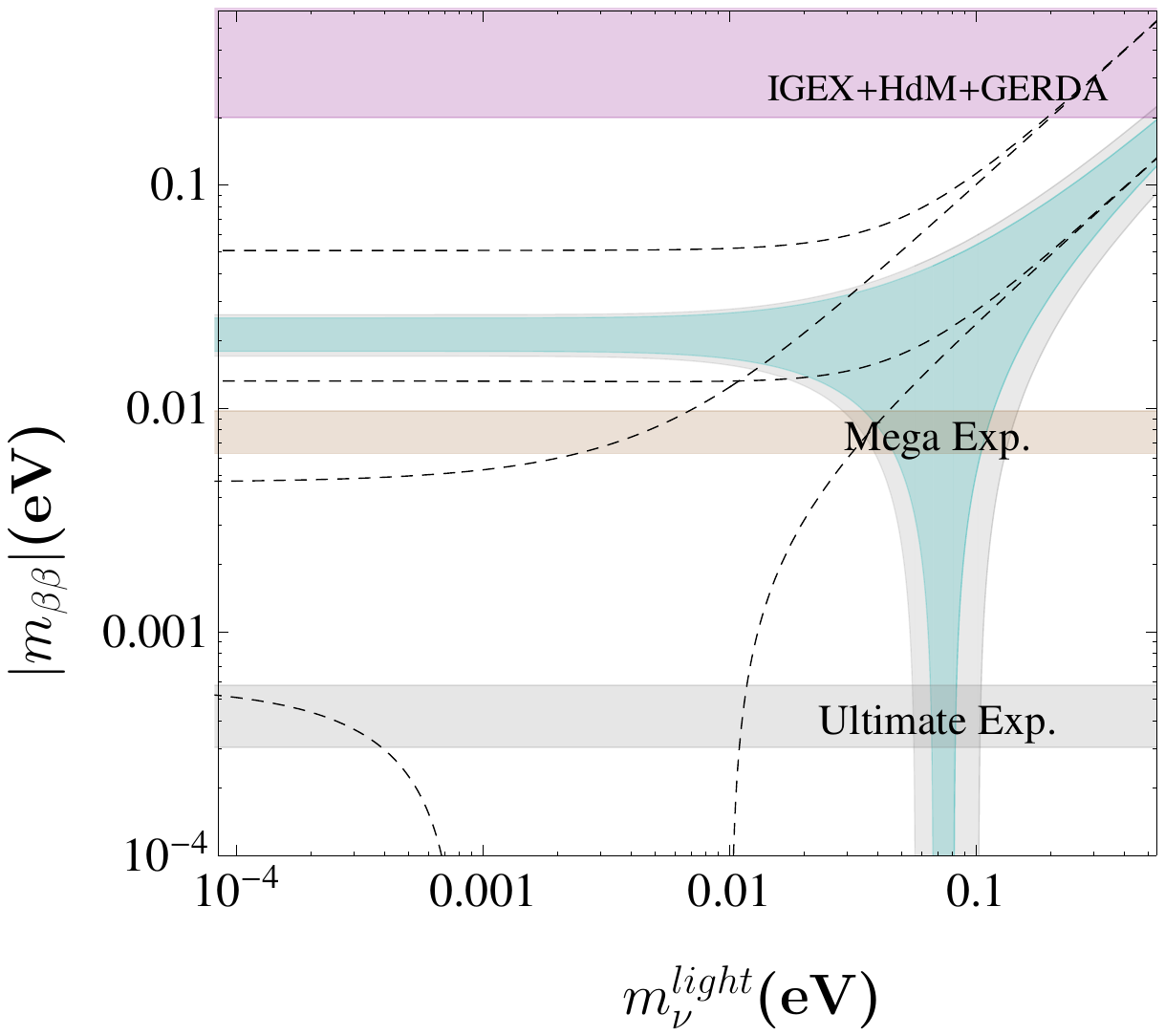}}
 \subfigure
 {\includegraphics[width=5cm,bb= 128 395 476 705]{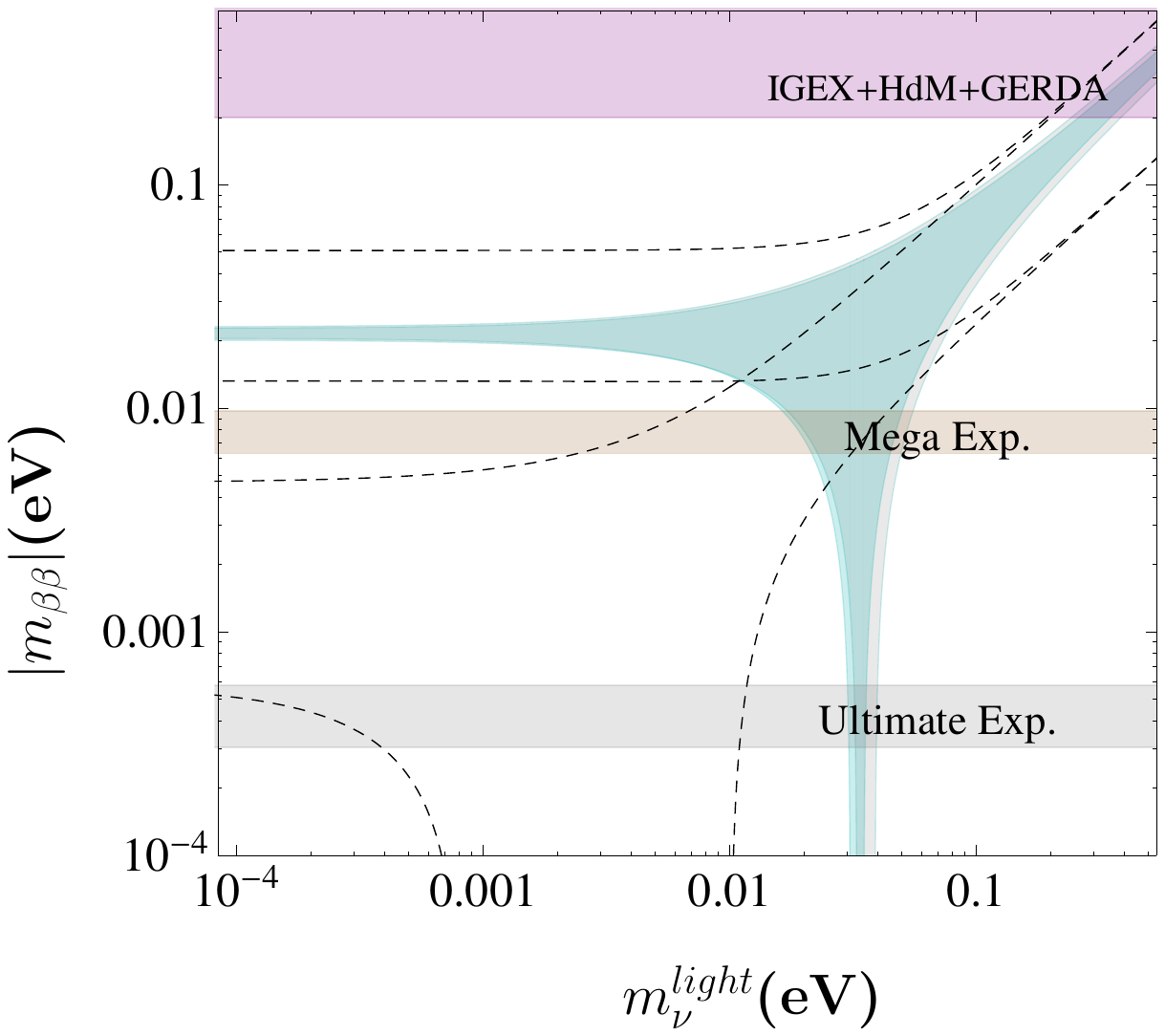}}
 \subfigure
 {\includegraphics[width=5cm,bb= 128 395 476 705]{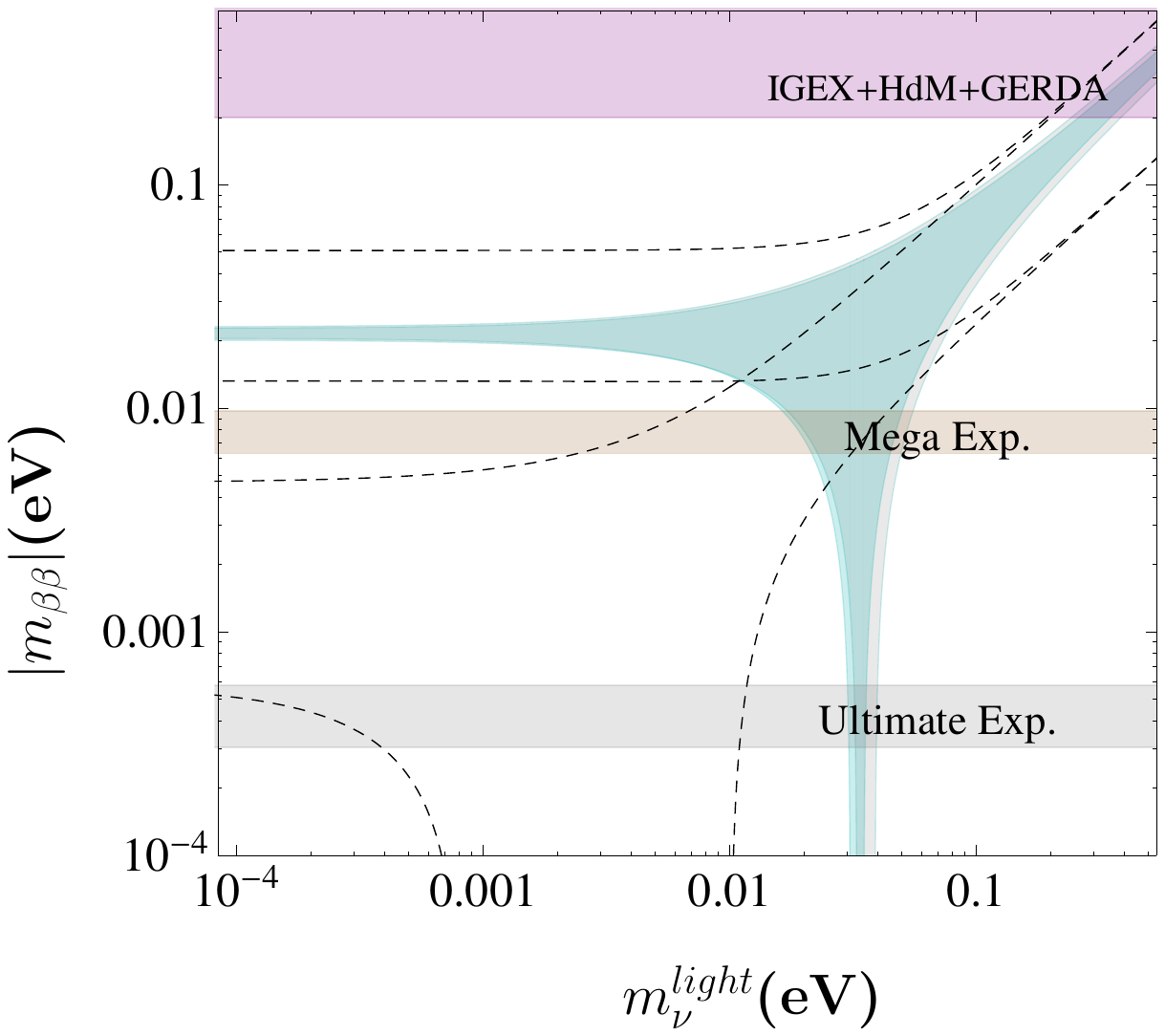}}
     \vspace*{-10pt}\caption{\label{fig:meffpseudosterile}
     Plots for \meff vs  $m_{\nu}^{light}$ for the NH case in the presence of one extra sterile neutrino
    in the case in which respectively, from
    the left to the right,    $\chi_1$ or $\chi_2$ 
    or $\chi_3$ are Dirac fermions.
    The light (Gray) shaded area is the $3\sigma$ allowed range given by oscillation data while
    the darker (Blue) shaded area is obtained from the best fit values in \cite{Capozzi:2013csa}. 
    The  dashed contour 
    corresponds to the predictions for the 3$\nu$ case.}
 \end{figure}
 \end{center}

As in the previous section we analyze first of all the cases in which one of 
the lightest states $\chi_1$, $\chi_2$ or $\chi_3$ is a Dirac neutrino. The results are shown in 
Fig. (\ref{fig:meffpseudosterile})
and  Fig. (\ref{fig:meffpseudosterileIH}) for NH and IH respectively. In both plots we consider a
3$\sigma$ uncertainty in the oscillation parameters which corresponds to the light shaded region while the 
colour shaded area determines the allowed range for the best fit values of ref. \cite{Capozzi:2013csa}.
As it is clear from the plots strong cancellations could be at work in  \meff 
for most of the cases both in the NH and in IH. In the NH case the intervals for the lightest mass where a complete
cancellation occurs are in the region $m_\nu^{light}\equiv m_1>  0.01$ eV. In the IH scheme instead the 
cancellation can be realized
 for smaller, or even zero, $m_\nu^{light}\equiv m_3$. There is only one case, 
 e.g. when $\chi_2$ is the Dirac particle, where for any value of $m_3$
  $\meff\gtrsim 5$ meV. In this case \meff has a similar behaviour with respect to the standard IH 
three neutrino case.\\
More interestingly if a couple of the lightest states are Dirac fermions than of course
in some cases the relative cancellations among the different terms are less effective
so in the NH with Dirac $(\chi_1,\chi_2)$ and in IH in the cases with $(\chi_1,\chi_2)$ and $(\chi_2,\chi_3)$
\meff could well be test in the next generation of experiments like GERDA-II, CUORE, etc.
This is depicted in Fig. (\ref{fig:meffpseudosterile2}) and  (\ref{fig:meffpseudosterile2IH}).

\onecolumngrid


\vspace*{-1cm}

 \begin{center}
 \begin{figure}[h!]
\subfigure
{\includegraphics[width=5cm,bb= 128 395 476 705]{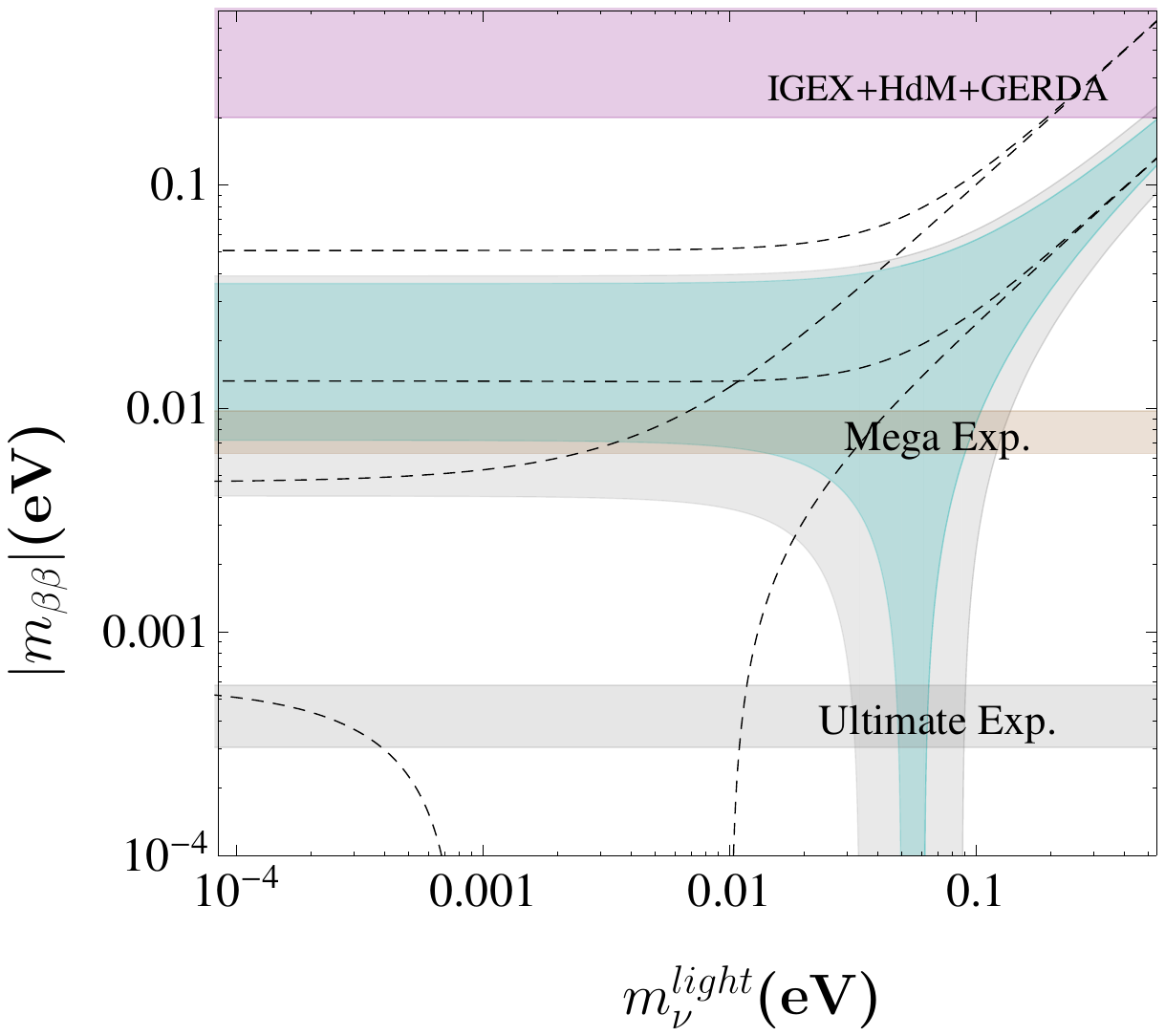}}
\subfigure
{\includegraphics[width=5cm,bb= 128 395 476 705]{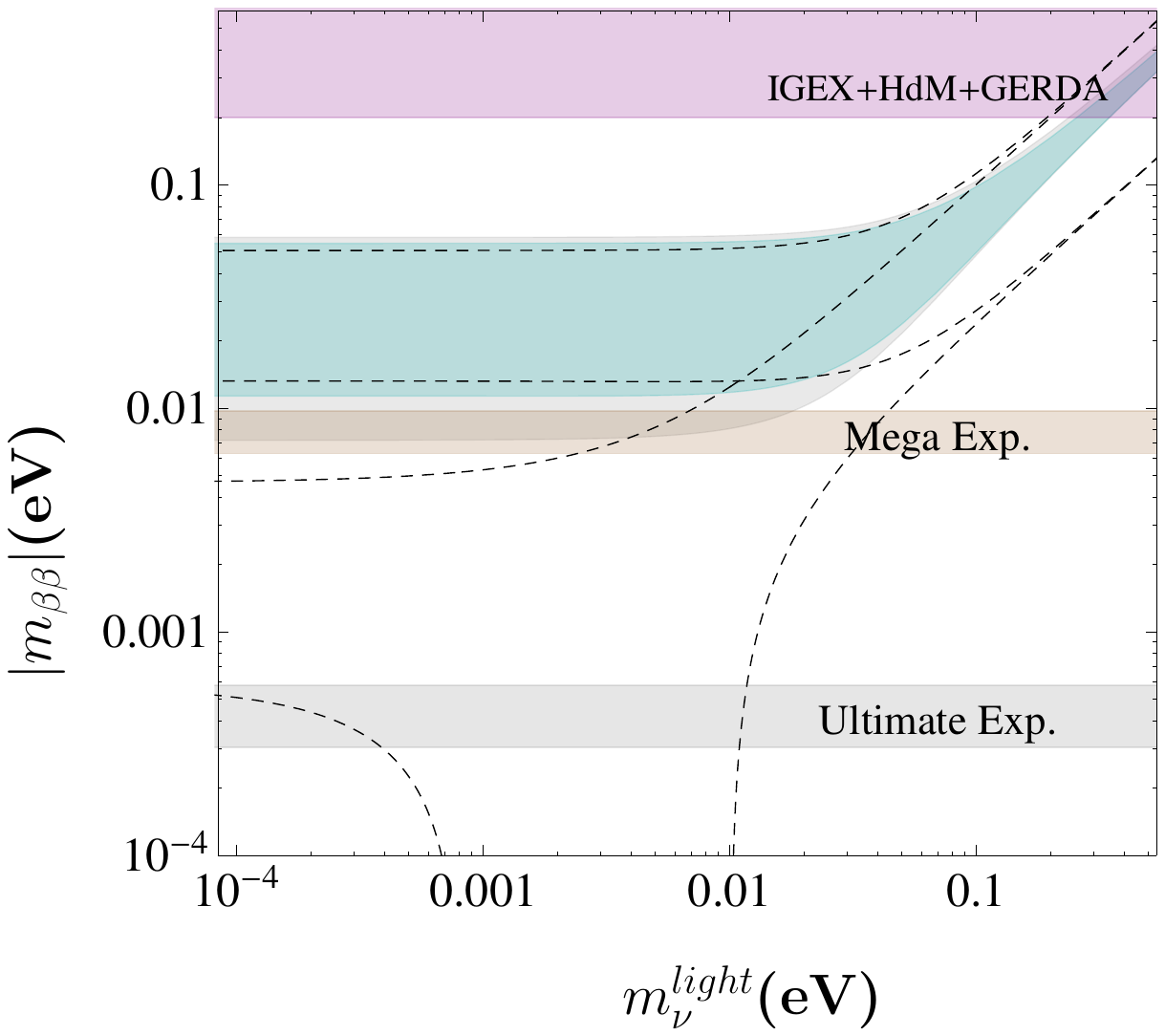}}
\subfigure
{\includegraphics[width=5cm,bb= 128 395 476 705]{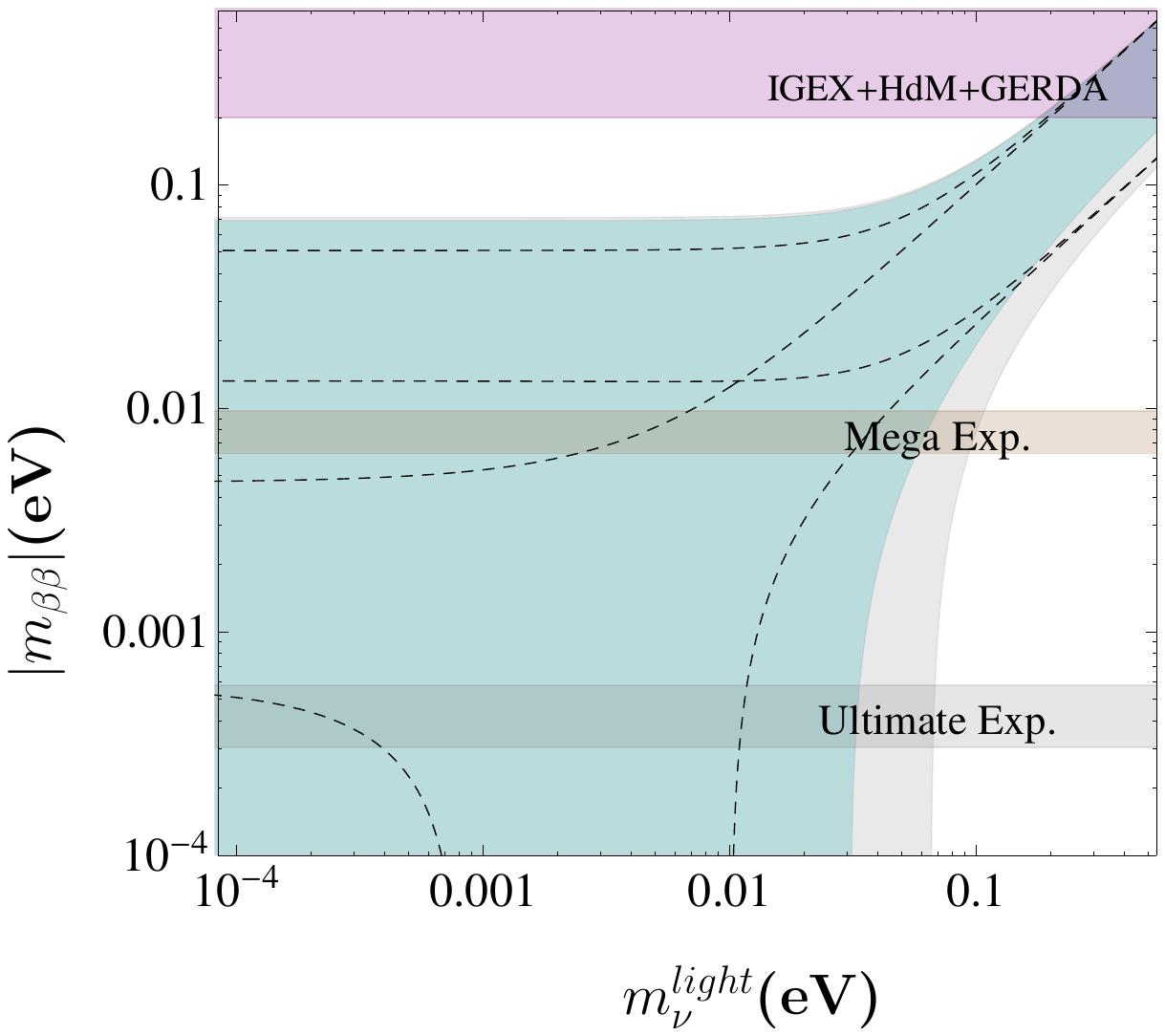}}
     \vspace*{-10pt}\caption{\label{fig:meffpseudosterileIH}
     Plots for \meff vs $m_{\nu}^{light}$ for the IH case in the presence of one extra sterile neutrino
    in the case in which respectively, from
    the left to the right,    $\chi_1$ or $\chi_2$ 
    or $\chi_3$ are Dirac fermions.  The light and darker shaded areas are given as in Fig. 
    \ref{fig:meffpseudosterile}.}
\end{figure}
\vskip-1cm
\begin{figure}[h!]
 \subfigure
 {\includegraphics[width=5cm,bb= 128 395 476 705]{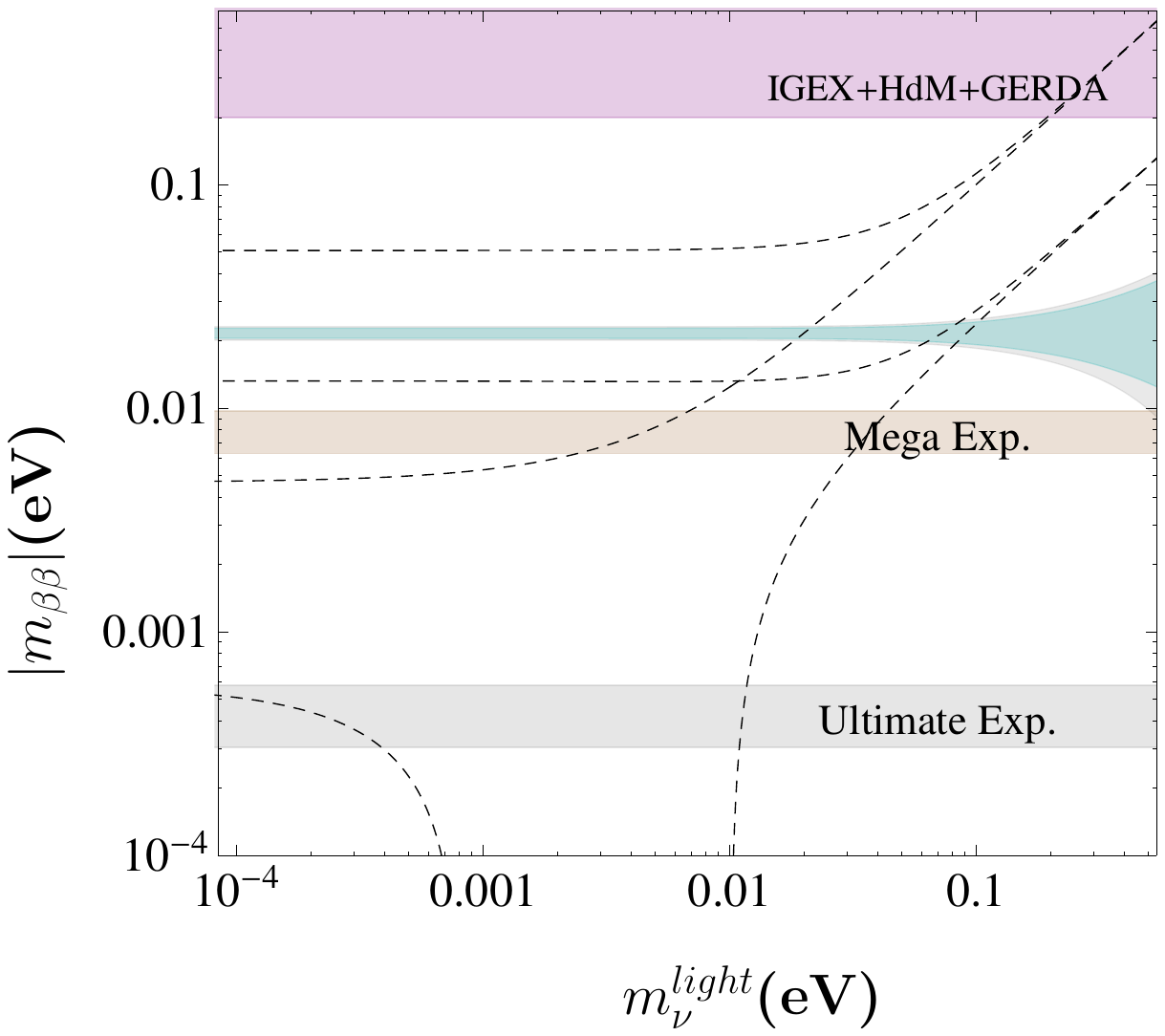}}
 \subfigure
 {\includegraphics[width=5cm,bb= 128 395 476 705]{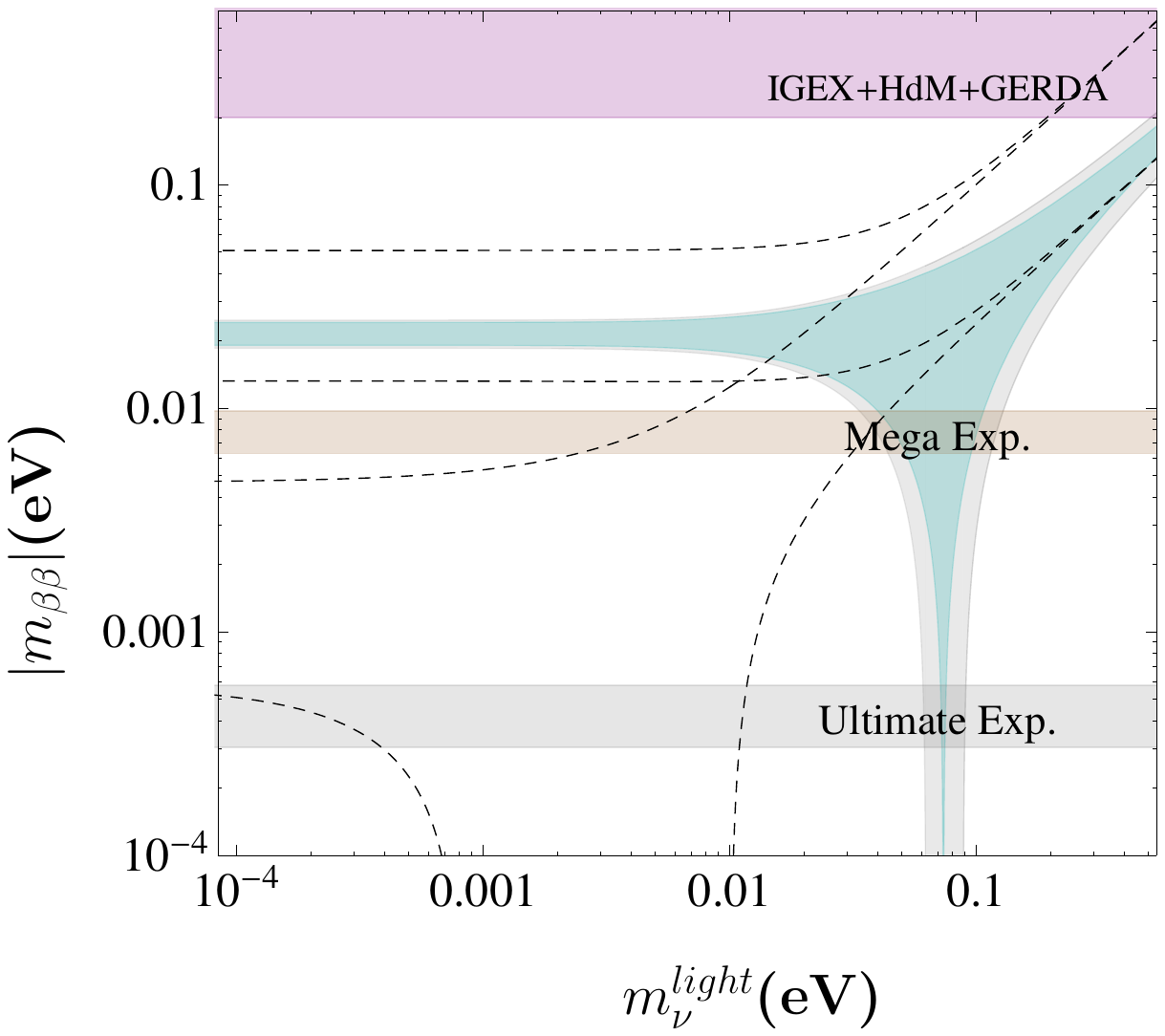}}
 \subfigure
 {\includegraphics[width=5cm,bb= 128 395 476 705]{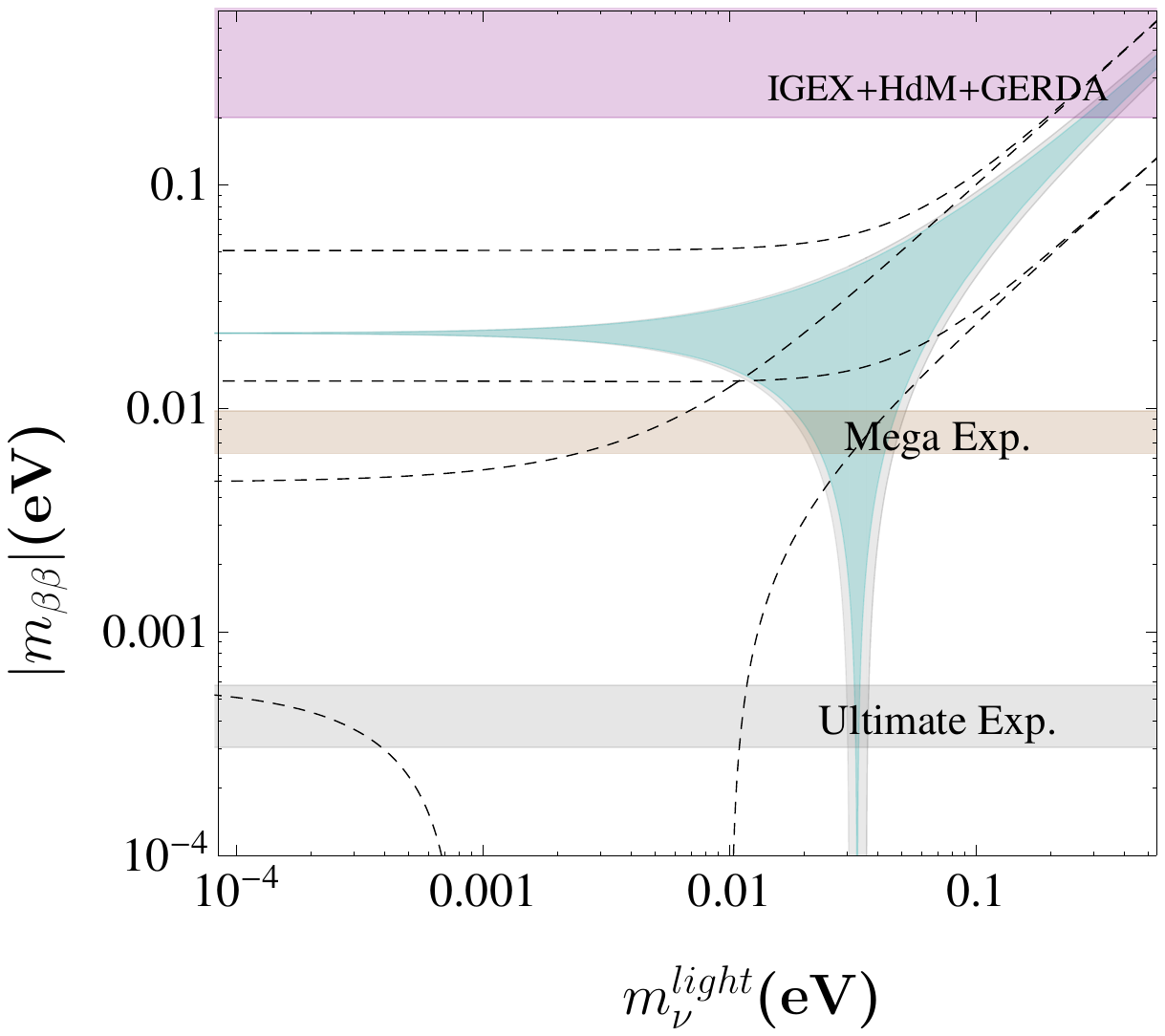}}
     \vspace*{-10pt}\caption{\label{fig:meffpseudosterile2}
     Plots for \meff vs $m_{\nu}^{light}$ for the NH case in the presence of one extra sterile neutrino state and
    in the case in which respectively, from
    the left to the right,    $(\chi_1,\chi_2)$ or $(\chi_1,\chi_3)$ 
    or $(\chi_2,\chi_3)$ are Dirac fermions. The light and darker shaded areas are given as in Fig. 
    \ref{fig:meffpseudosterile}.}
\end{figure}
\vskip-1cm
\begin{figure}[h!]
 \subfigure
 {\includegraphics[width=5cm,bb= 128 395 476 705]{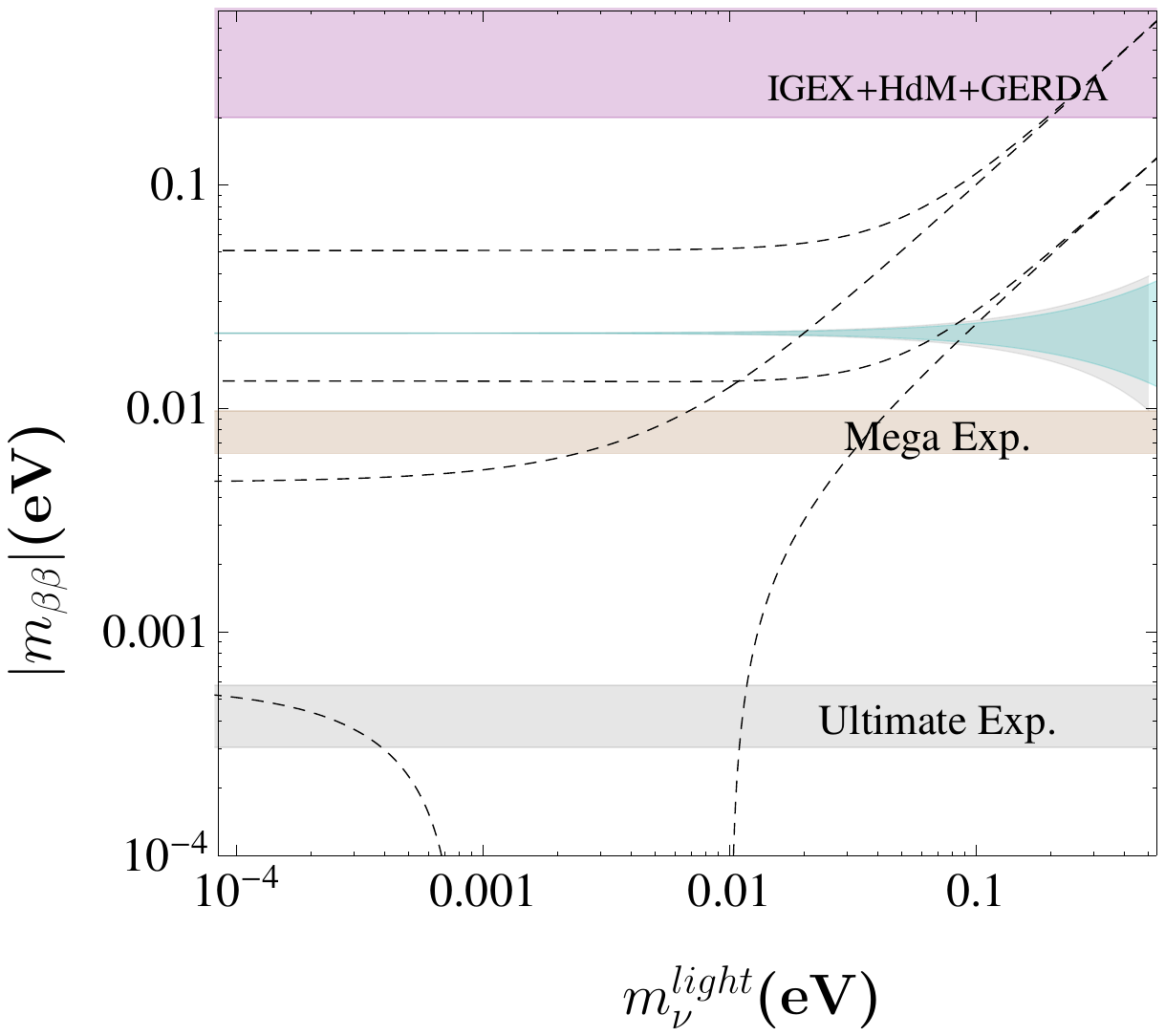}}
 \subfigure
 {\includegraphics[width=5cm,bb= 128 395 476 705]{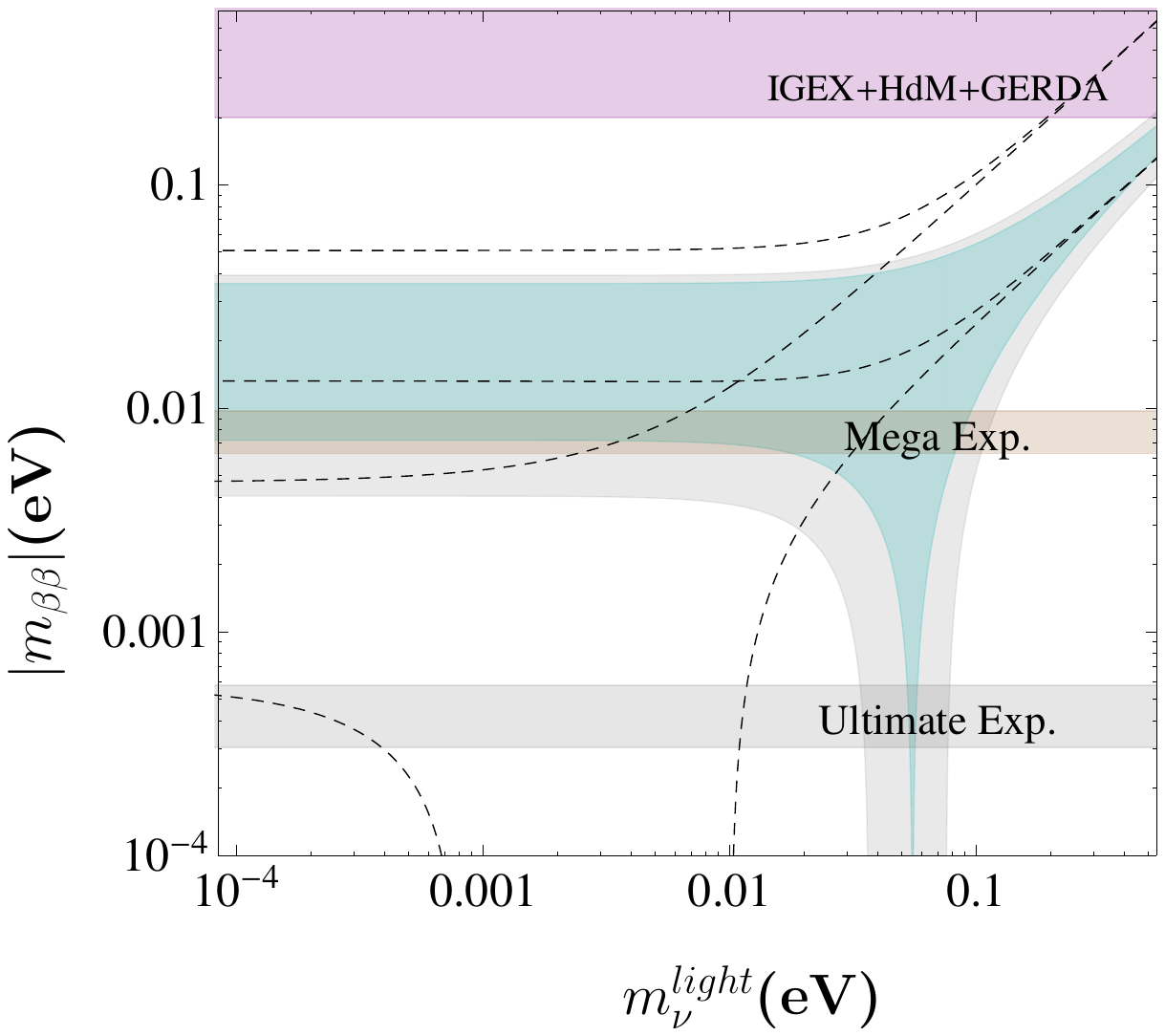}}
 \subfigure
 {\includegraphics[width=5cm,bb= 128 395 476 705]{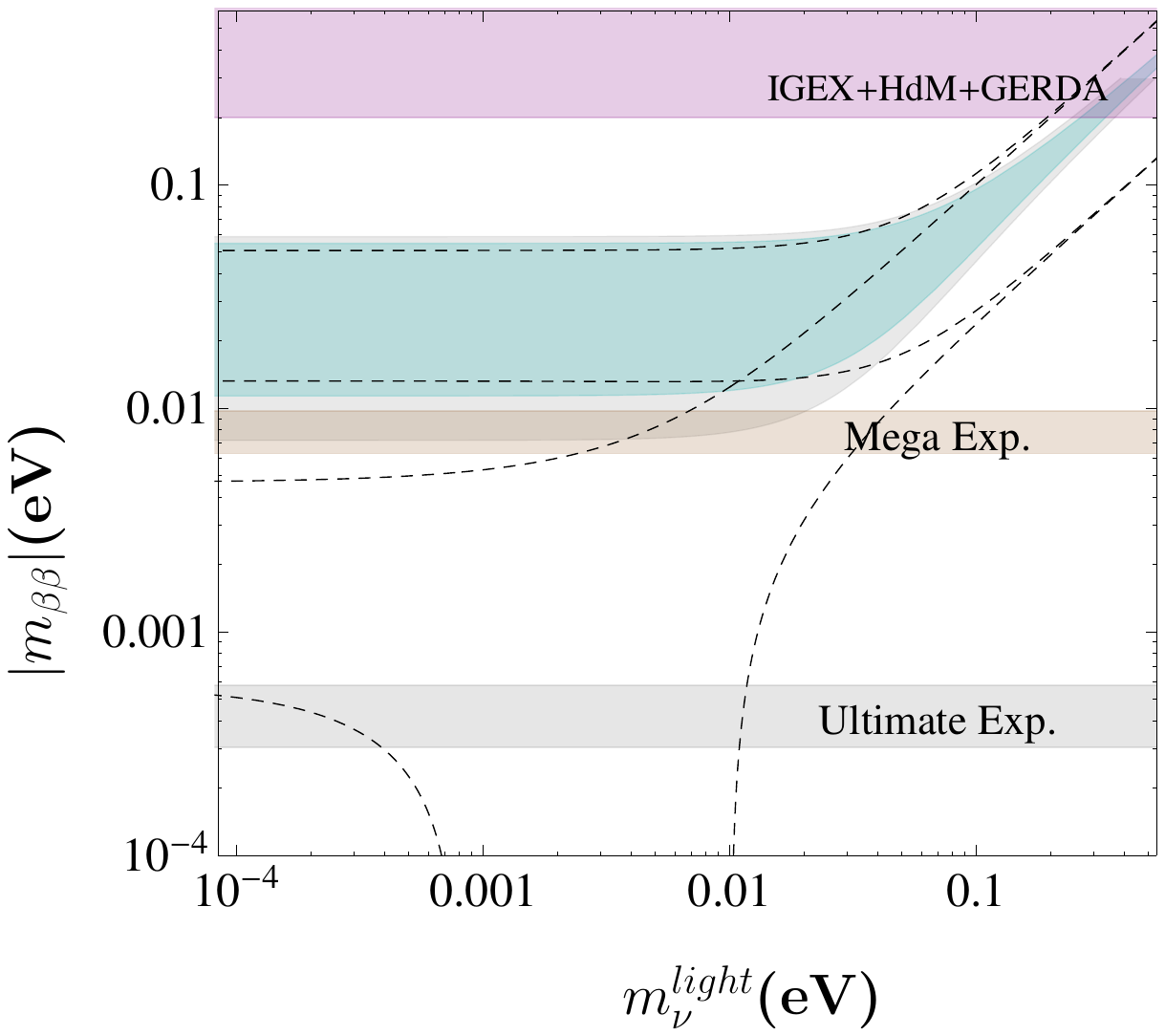}}
     \vspace*{-10pt}\caption{\label{fig:meffpseudosterile2IH}
      Plots for \meff vs $m_{\nu}^{light}$ for the IH case in the presence of one extra sterile neutrino state and
    in the case in which respectively, from
    the left to the right,    $(\chi_1,\chi_2)$ or $(\chi_1,\chi_3)$ 
    or $(\chi_2,\chi_3)$ are Dirac fermions.  
    The light and darker shaded areas are given as in Fig. \ref{fig:meffpseudosterile}.}
\end{figure}
\end{center}

\newpage
\twocolumngrid
\section{Discussion}

The current and near term neutrinoless double beta decay (\betabeta-decay) experiments
will be able to explore the region for $\meff\gtrsim 0.01$ eV. 
Hopefully with experiments such as GERDA-II, CUORE, nEXO etc. most of this region will be
investigated but if the decay is not observed new experimental efforts will be required.
From the theoretical point of view, it is also necessary to understand which  phenomenological frameworks
could be tested in the next generation of \betabeta-decay experiments and also with the future {\it mega}-
and {\it ultimate}-generations. 
In particular, it makes sense to investigate which scenarios for the NH spectrum can be experimentally reachable in the {\it mega-} and 
{\it ultimate}-generations, since in the standard three Majorana neutrino framework a complete cancellation can occur.
\\
In this paper we have analyzed the case in which one or two active neutrinos
are Dirac particles for both NH and IH. We found  that due to 
narrow $3\sigma$ range for $\theta_{13}$ most of the cases we analyzed with one 
Dirac neutrino, could be falsifiable in the \betabeta-decay {\it mega}-generation
since a complete cancellation not always occurs.
In fact for the NH case there is no cancellation when the neutrino $\chi_1$ is Dirac, 
see the plot on the left of Fig. (\ref{fig:meffpseudo}).
On the opposite, if  a complete cancellation is realized, this occurs 
in a region where the lightest state is almost massless, see the plot in the center and  on the right of
Fig. (\ref{fig:meffpseudo}). 
In the case of two Dirac neutrinos, 
a cancellation occurs only when the Dirac neutrinos 
are $(\chi_2,\chi_3)$, 
see the plot in the right side of Fig. (\ref{fig:meffpseudo2}).
\\
In the second part of the paper we have analyzed the allowed region for \betabeta-decay in the presence of pseudo-Dirac neutrinos
together with an extra sterile neutrino state with mass of the order of 1 eV,
as hinted by a number of anomalies arising in short baseline neutrino oscillation experiments.
In this case if one of the active state is Dirac then in all the cases  we analyzed but one,
strong cancellations can  occur.
However, it is notable that for NH, these scenarios are considerably different from  the standard three-neutrino case
because the interval for the lightest neutrino mass 
in which the \meff is zero is different. 
If instead two of the active states are Dirac fermions then the situation is more favorable
and some cases are in the range of the next generation experiments.
In summary, it makes sense from a theoretical perspective to analyze  different scenarios
for the nature of massive neutrinos that could be tested through  terrestrial experiments. This
means not only  neutrinos as pure Dirac or pure Majorana particles but also the possibility that 
both natures coexist, even if from the theoretical point of view is less appealing.

\section{Acknowledgments}
The authors fully acknowledge interesting correspondence with R. Brugnera, 
S. Pastor and E. Giusarma. A.M. acknowledges MIUR (Italy)  for financial support under
the program ``Futuro in Ricerca 2010'', (RBFR10O36O).

\section{Appendix}
\begin{center}
\begin{figure}[h!]
 \subfigure
 {\includegraphics[width=5cm,bb= 128 395 476 705]{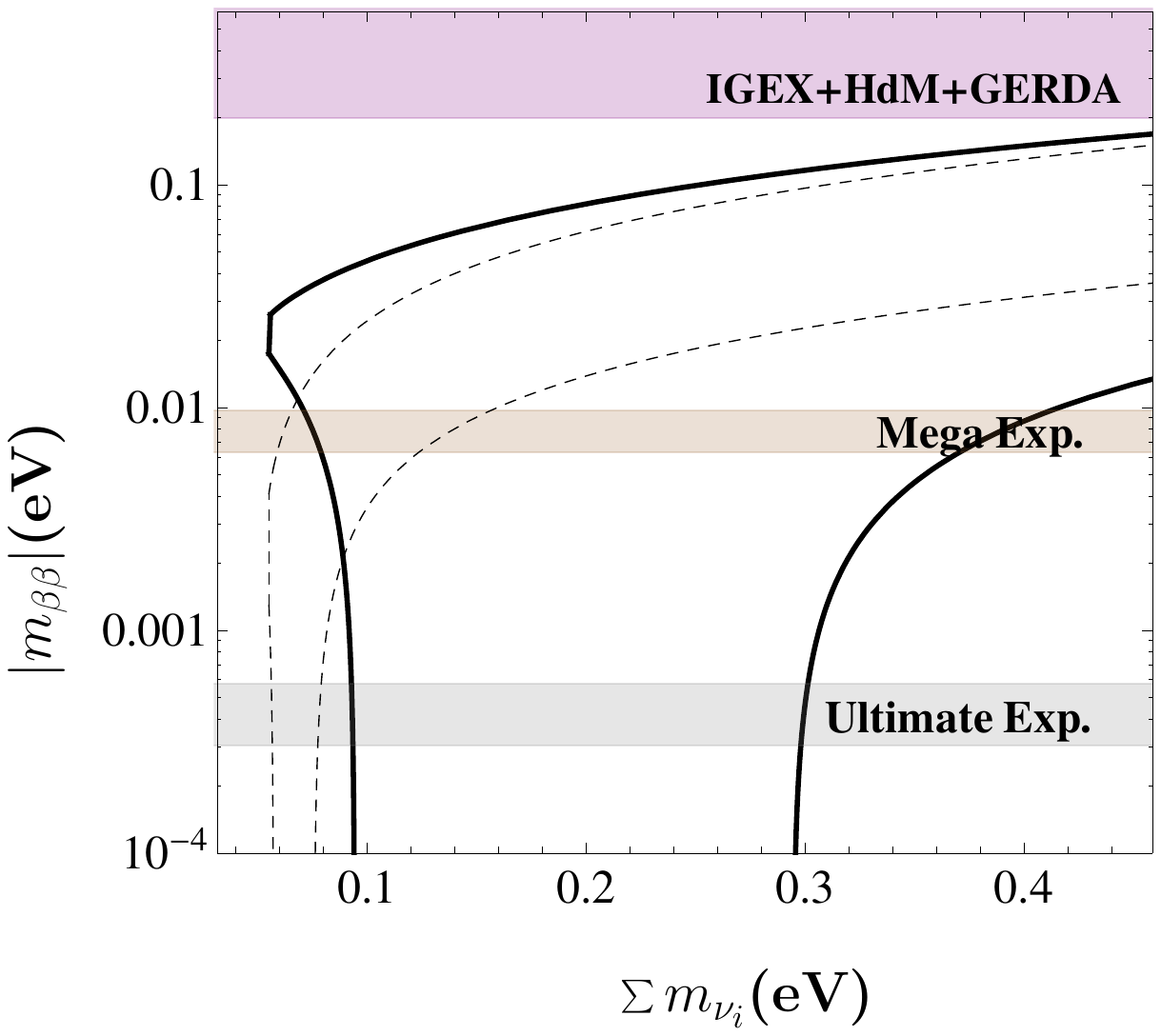}}
 \hspace{1cm}\subfigure
 {\includegraphics[width=5cm,bb= 128 395 476 705]{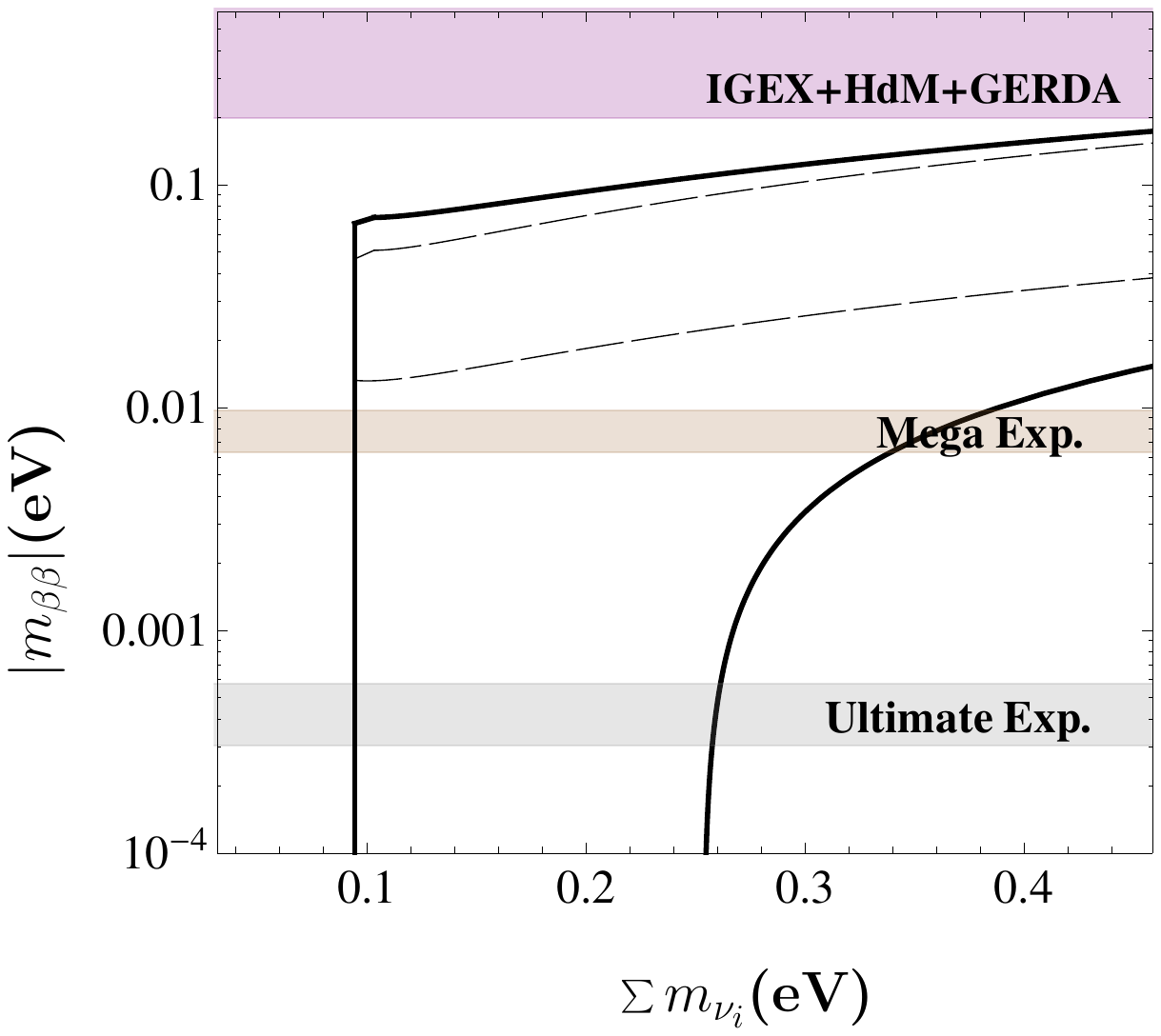}}
     \vspace*{-10pt}\caption{\label{fig:meffsterileapp}Plots for \meff vs $\sumass$ 
    in the 3+1 scenario i.e. with one extra sterile state with $\Delta m^2_{SBL}=0.93 eV^2$ and 
    $\sin\theta_{14}=0.15$ for NH (left panel) and IH (right panel). The dotted contours 
    define the allowed range for \meff in the $3\nu$
    framework.}
    \end{figure}
\end{center}

\begin{center}
\begin{figure}[h!]
 \subfigure
 {\includegraphics[width=5cm,bb= 128 395 476 705]{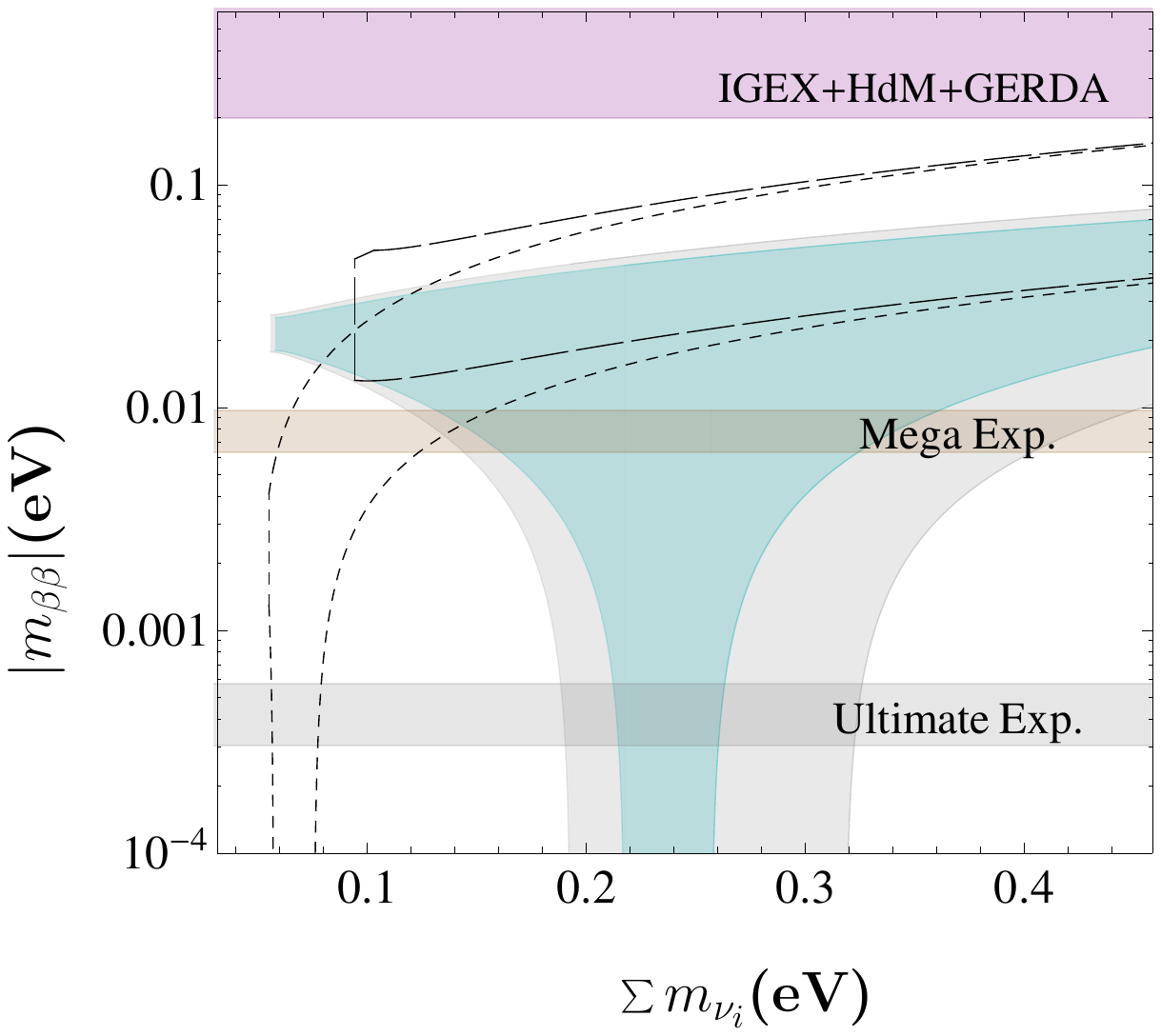}}
 \subfigure
 {\includegraphics[width=5cm,bb= 128 395 476 705]{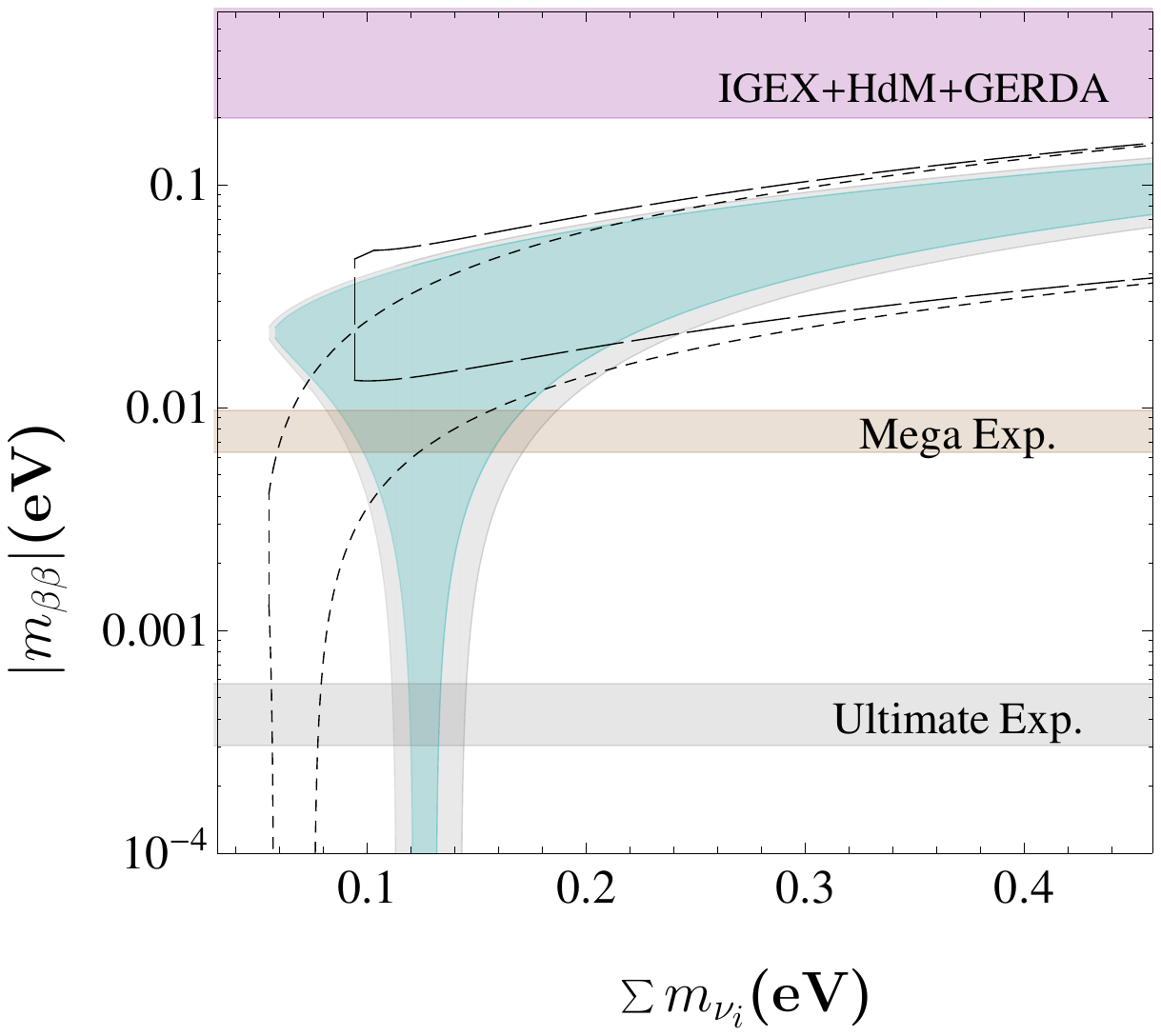}}
 \subfigure
 {\includegraphics[width=5cm,bb= 128 395 476 705]{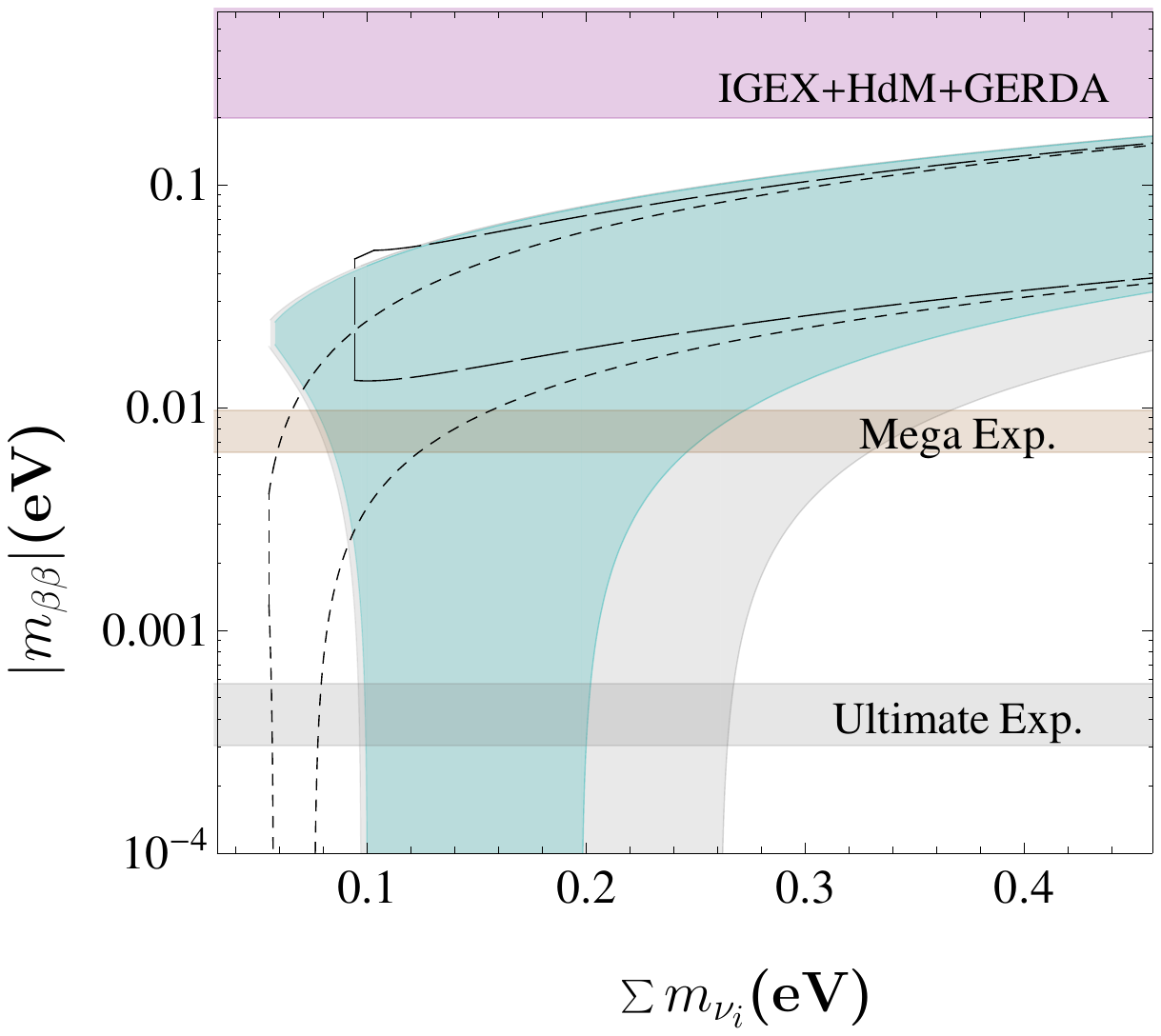}}
     \vspace*{-10pt}\caption{\label{fig:meffpseudosterileapp}
     Plots for \meff vs $\sumass$ in the 3+1 scenario for the NH case in the case in which respectively, from
    the left to the right,    $\chi_1$ or $\chi_2$ 
    or $\chi_3$ are Dirac fermions. The light (Gray) shaded area is the $3\sigma$ allowed range given by oscillation data while
    the darker (Blue) shaded area is obtained from the best fit values in \cite{Capozzi:2013csa}. The short (long) 
    dashed contours
    correspond to the predictions for the 3$\nu$ case.}
\end{figure}
\end{center}

\begin{center}
\begin{figure}[h!]
\subfigure
{\includegraphics[width=5cm,bb= 128 395 476 705]{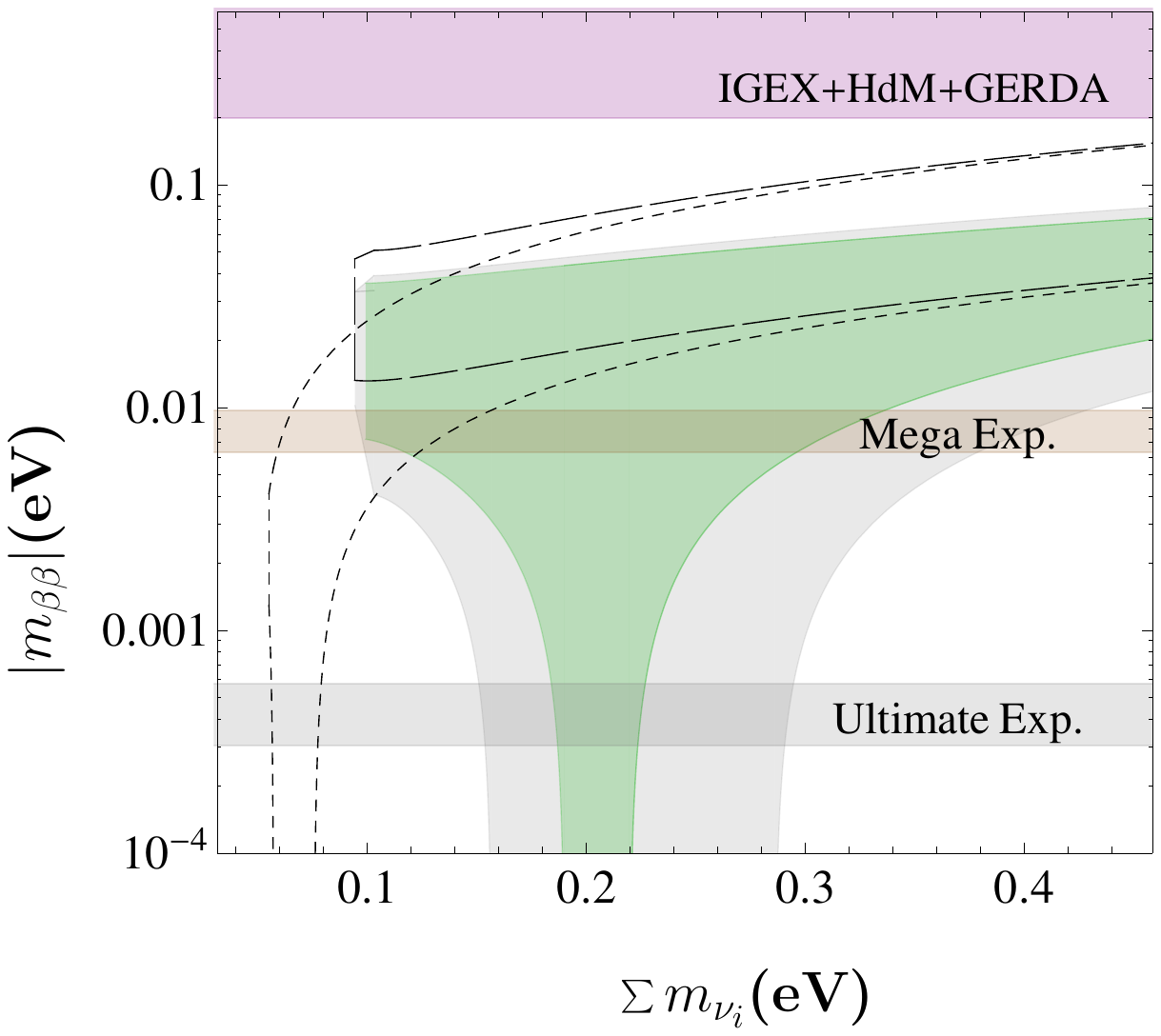}}
\subfigure
{\includegraphics[width=5cm,bb= 128 395 476 705]{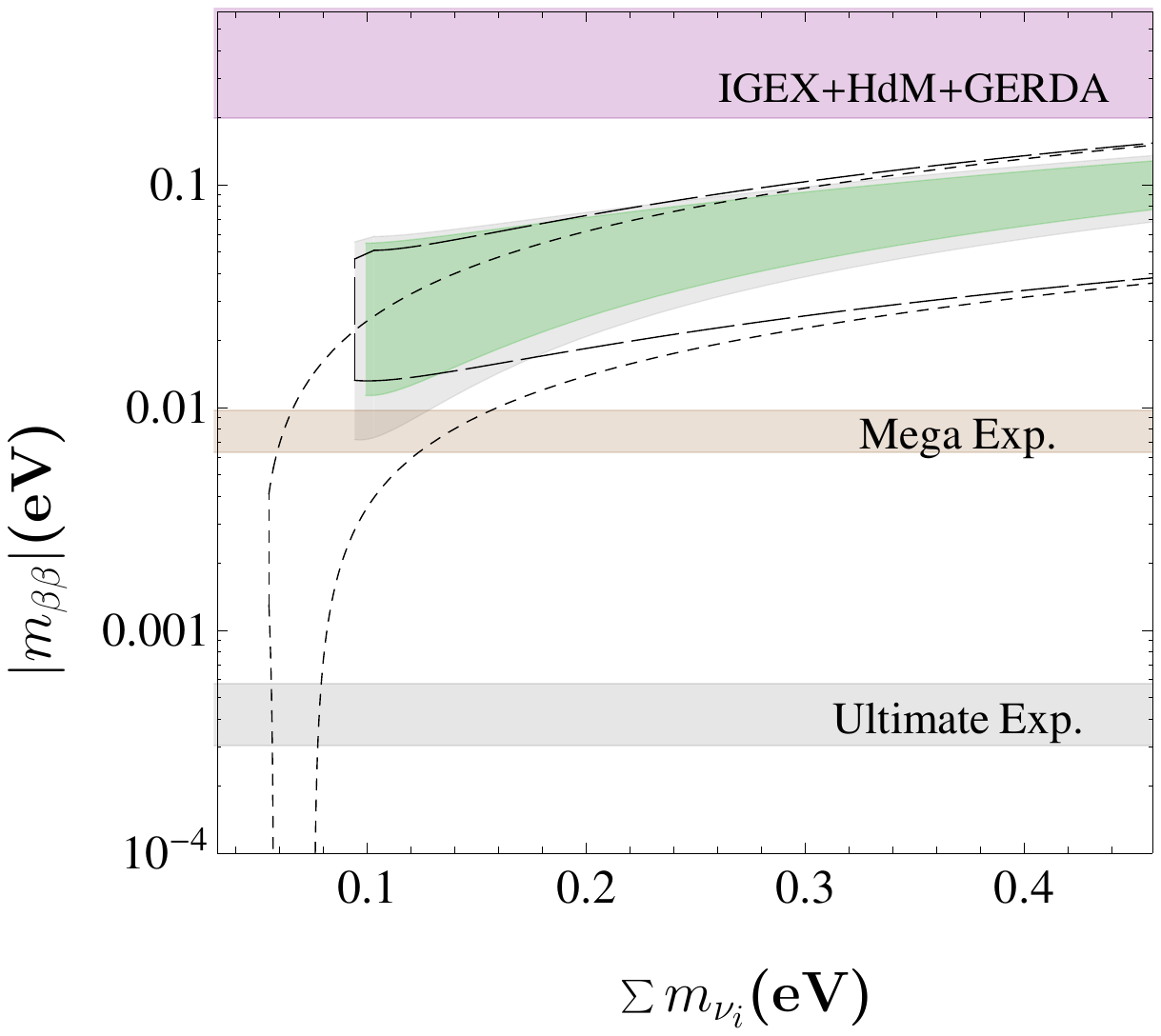}}
\subfigure
{\includegraphics[width=5cm,bb= 128 395 476 705]{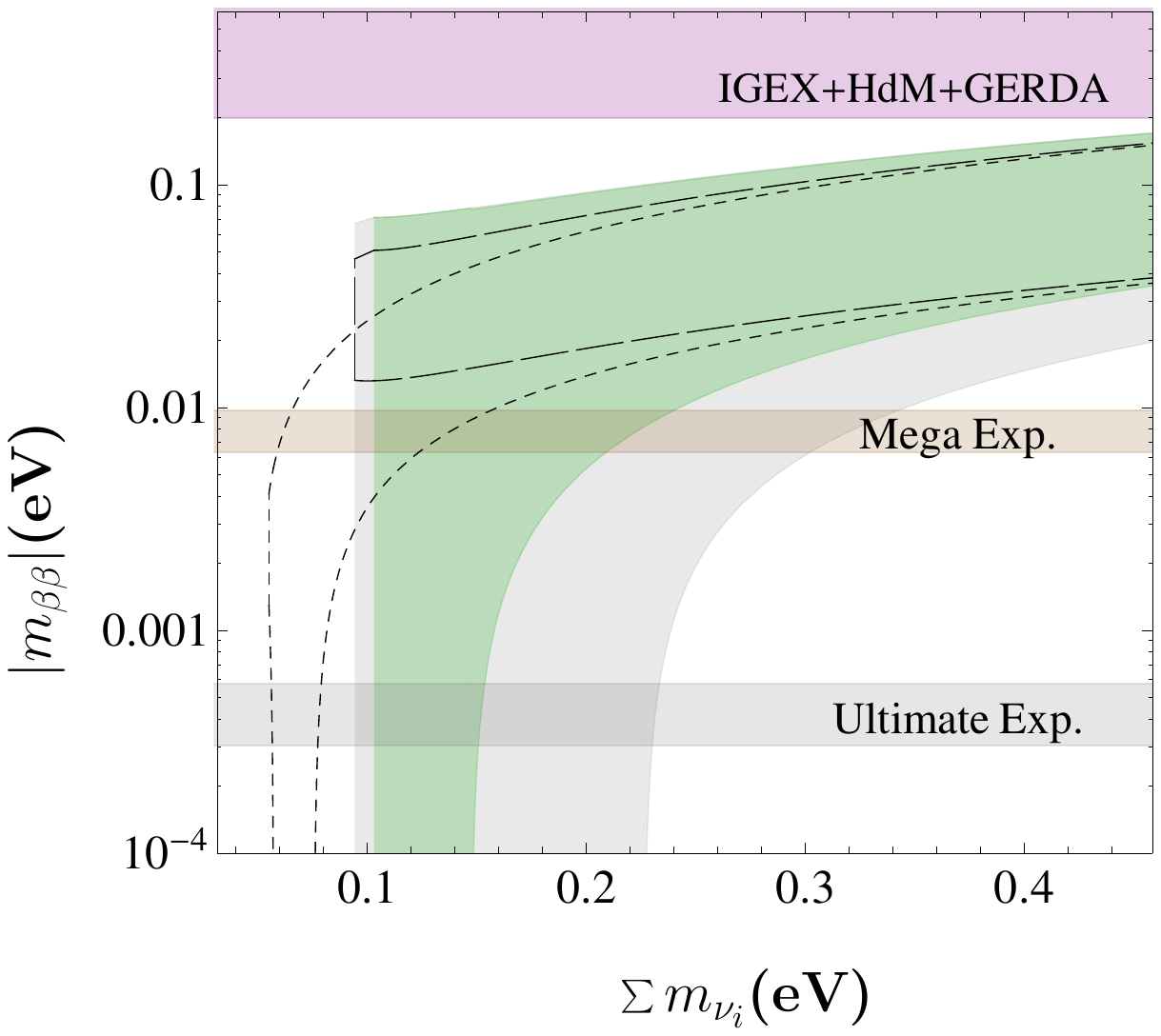}}
     \vspace*{-10pt}\caption{\label{fig:meffpseudosterileIHapp}
     Plots for \meff vs $\sumass$ for the IH case in the presence of one extra sterile neutrino
    in the case in which respectively, from
    the left to the right,    $\chi_1$ or $\chi_2$ 
    or $\chi_3$ are Dirac fermions.
    The light (Gray) and darker (Green) shaded areas indicate the allowed range for the
    $3\sigma$ and best fit values given in  \cite{Capozzi:2013csa}. }
\end{figure}
\end{center}

\begin{center}
\begin{figure}[h!]
 \subfigure
 {\includegraphics[width=5cm,bb= 128 395 476 705]{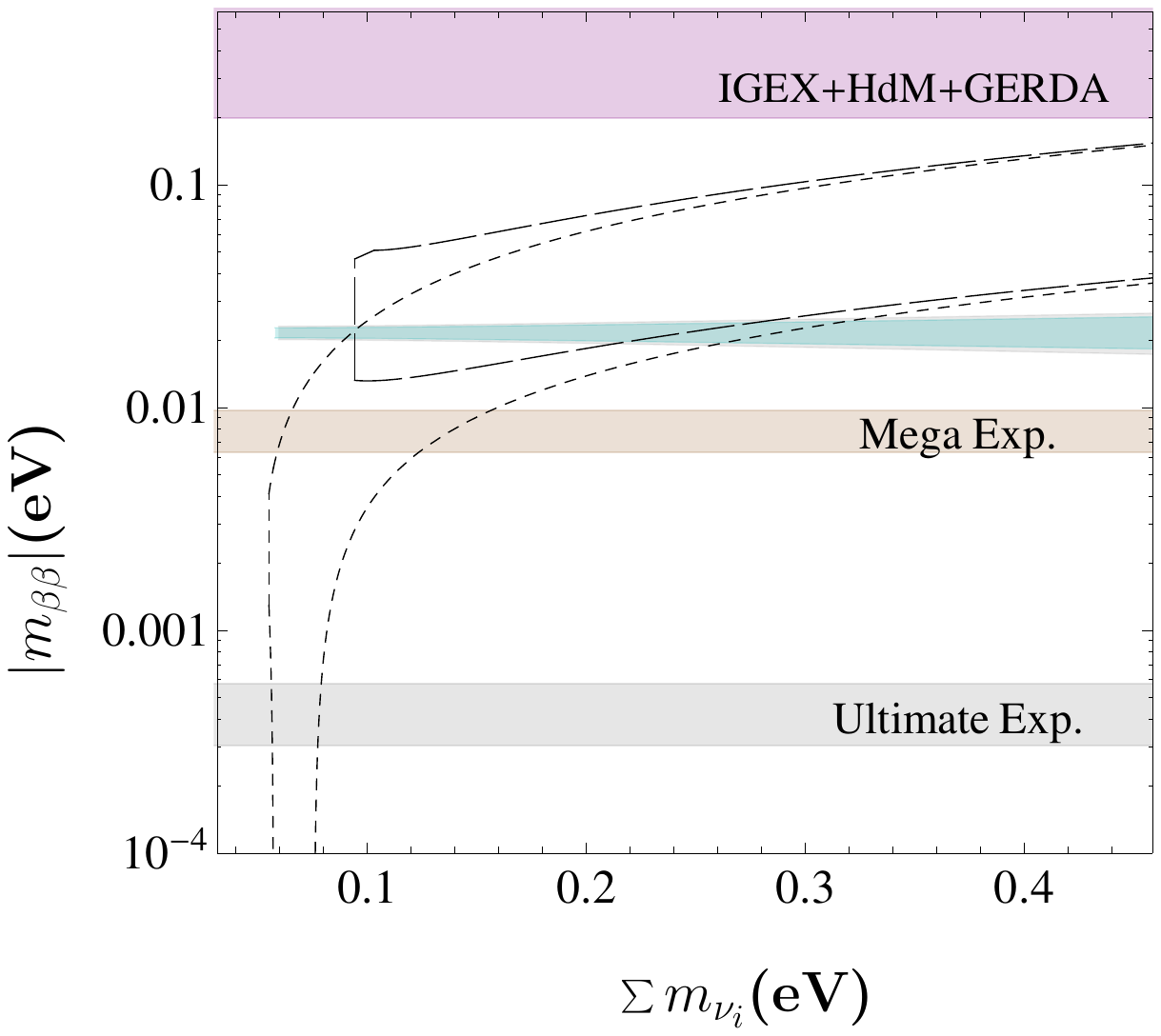}}
 \subfigure
 {\includegraphics[width=5cm,bb= 128 395 476 705]{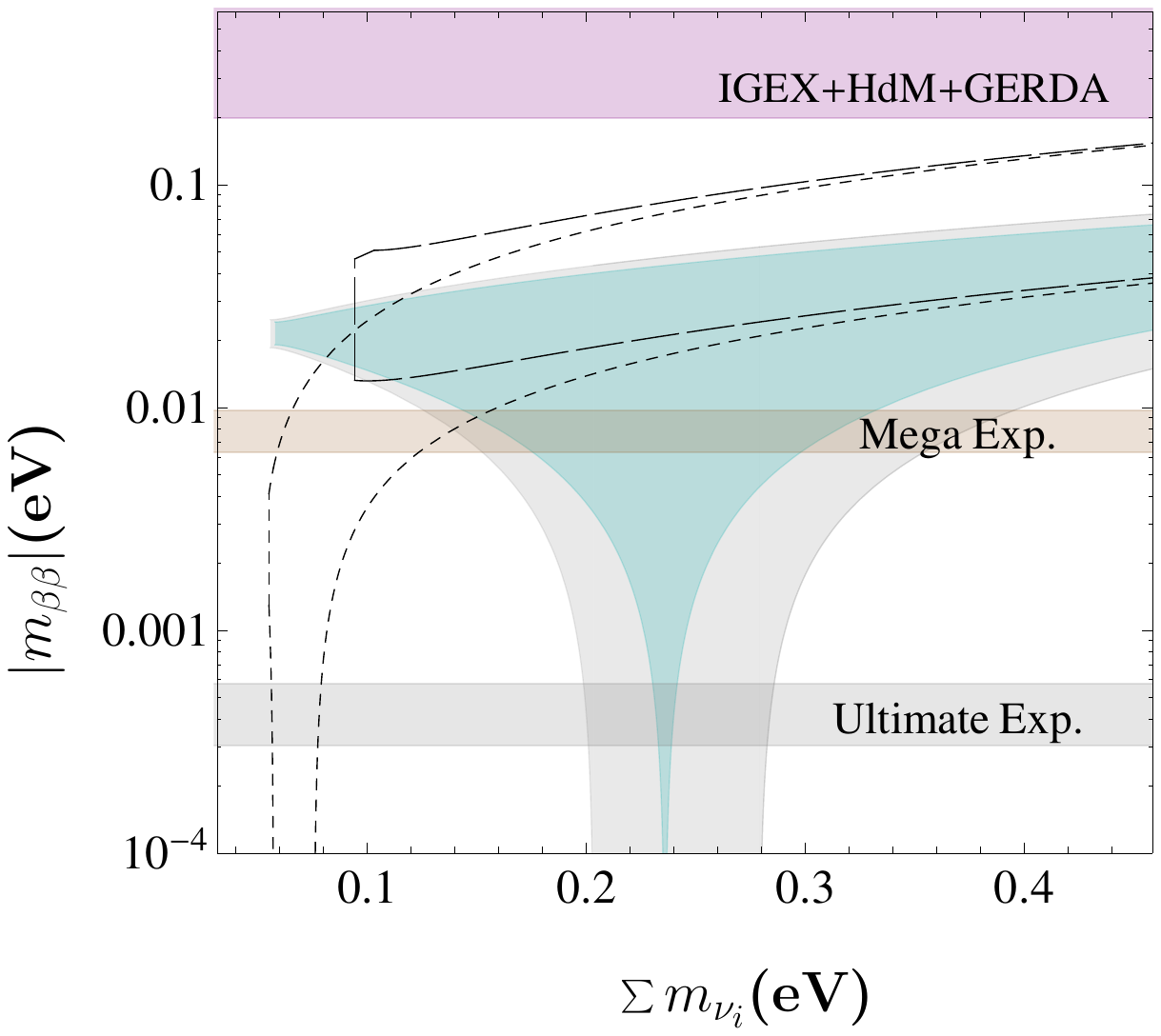}}
 \subfigure
 {\includegraphics[width=5cm,bb= 128 395 476 705]{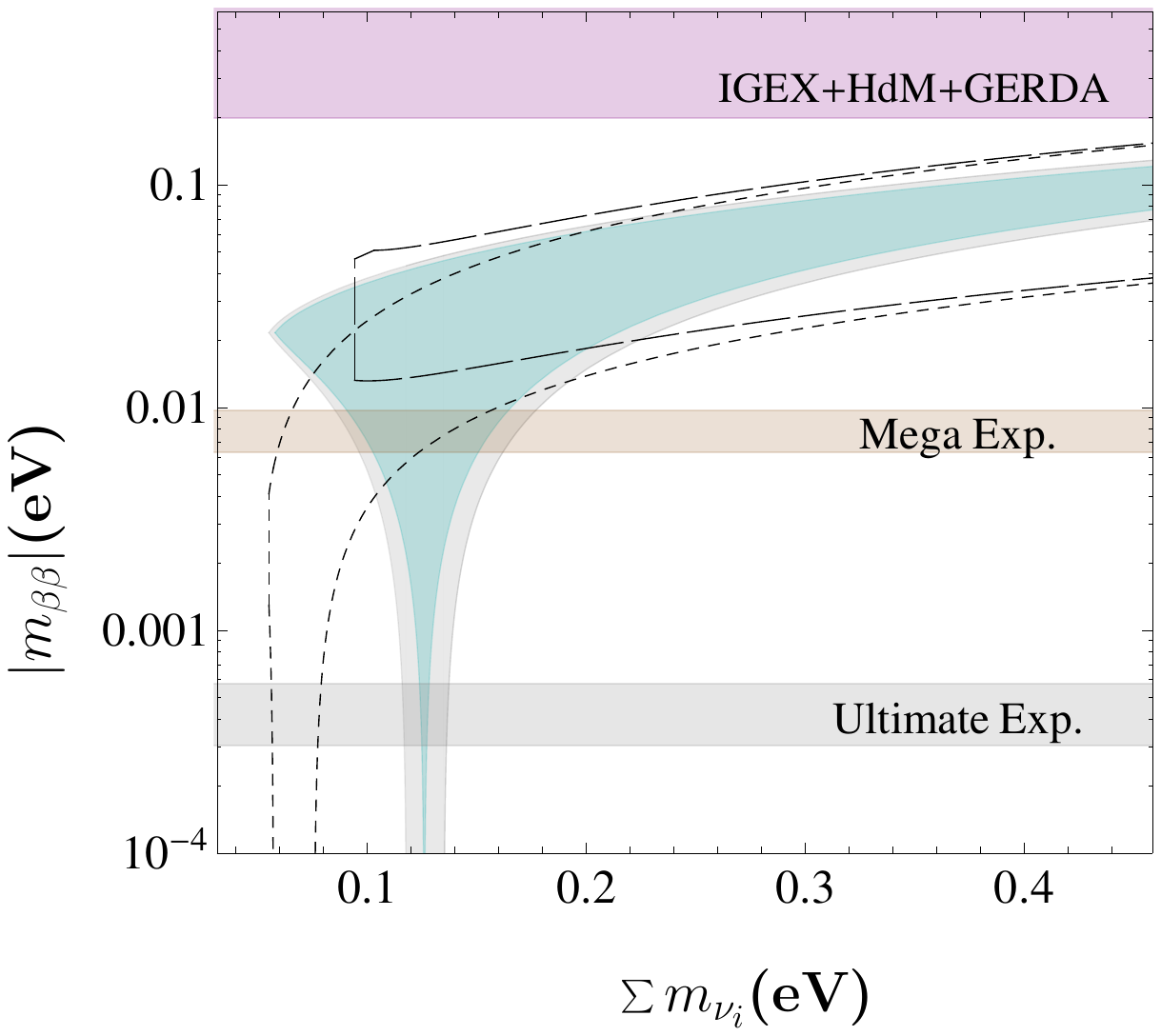}}
    \vspace*{-10pt} \caption{\label{fig:meffpseudosterile2app}
     Plots for \meff vs $\sumass$ for the NH case in the presence of one extra sterile neutrino state and
    in the case in which respectively, from
    the left to the right,    $(\chi_1,\chi_2)$ or $(\chi_1,\chi_3)$ 
    or $(\chi_2,\chi_3)$ are Dirac fermions.   The light  (Gray) 
    and darker (Blue) shaded areas are given as in Fig. \ref{fig:meffpseudosterileapp}.}
\end{figure}
\end{center}

\begin{center}
\begin{figure}[h!]
 \subfigure
 {\includegraphics[width=5cm,bb= 128 395 476 705]{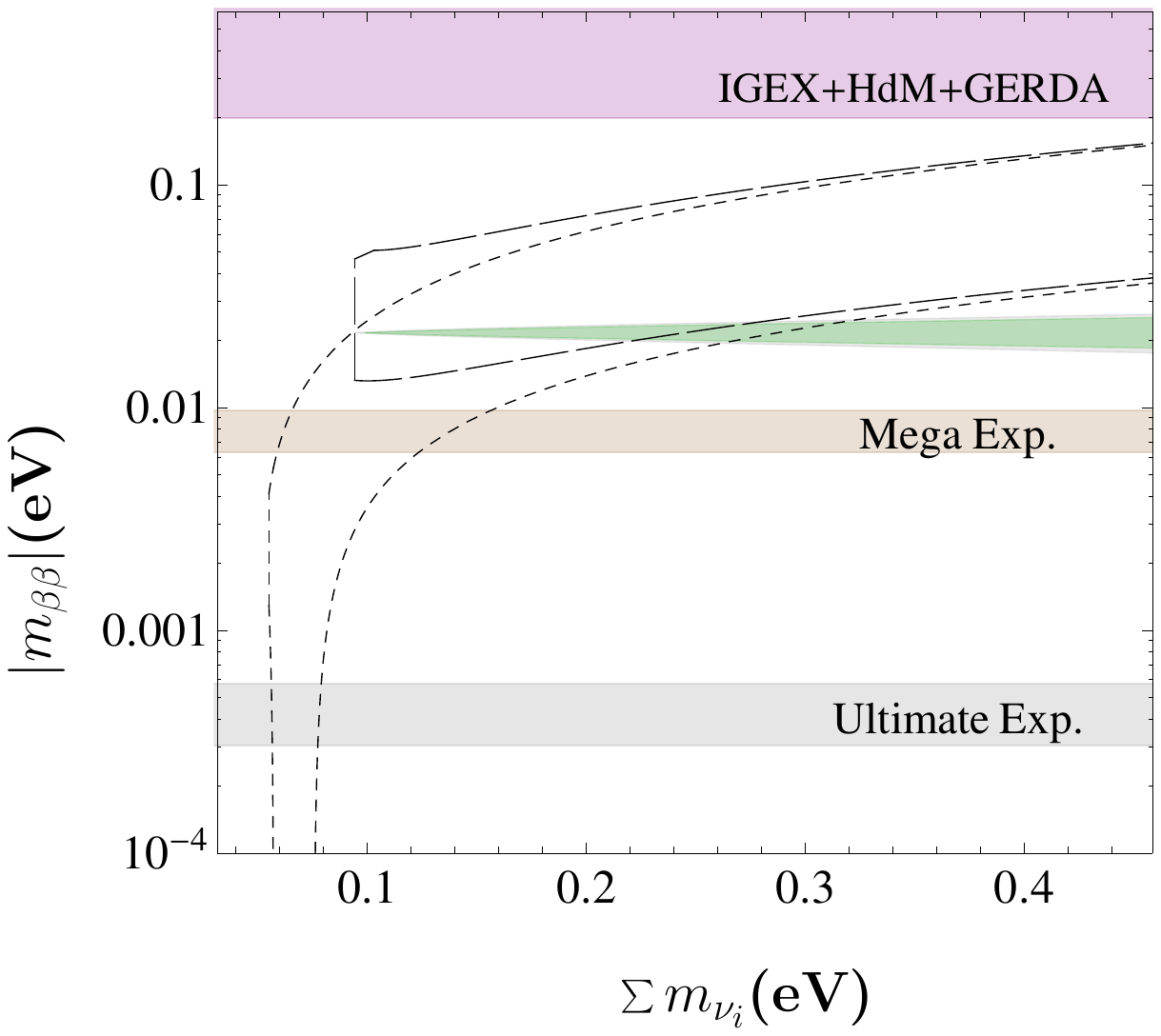}}
 \subfigure
 {\includegraphics[width=5cm,bb= 128 395 476 705]{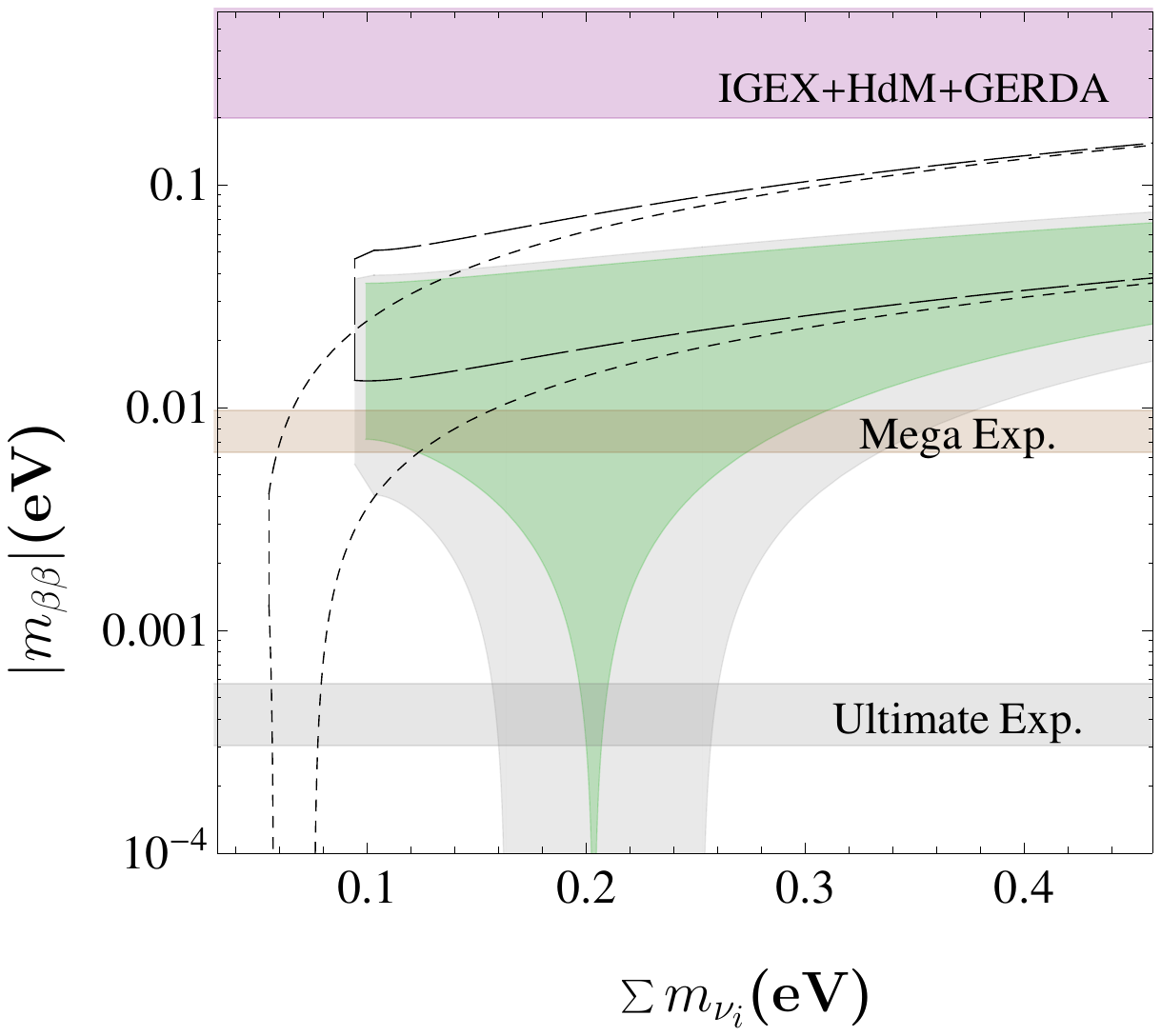}}
 \subfigure
 {\includegraphics[width=5cm,bb= 128 395 476 705]{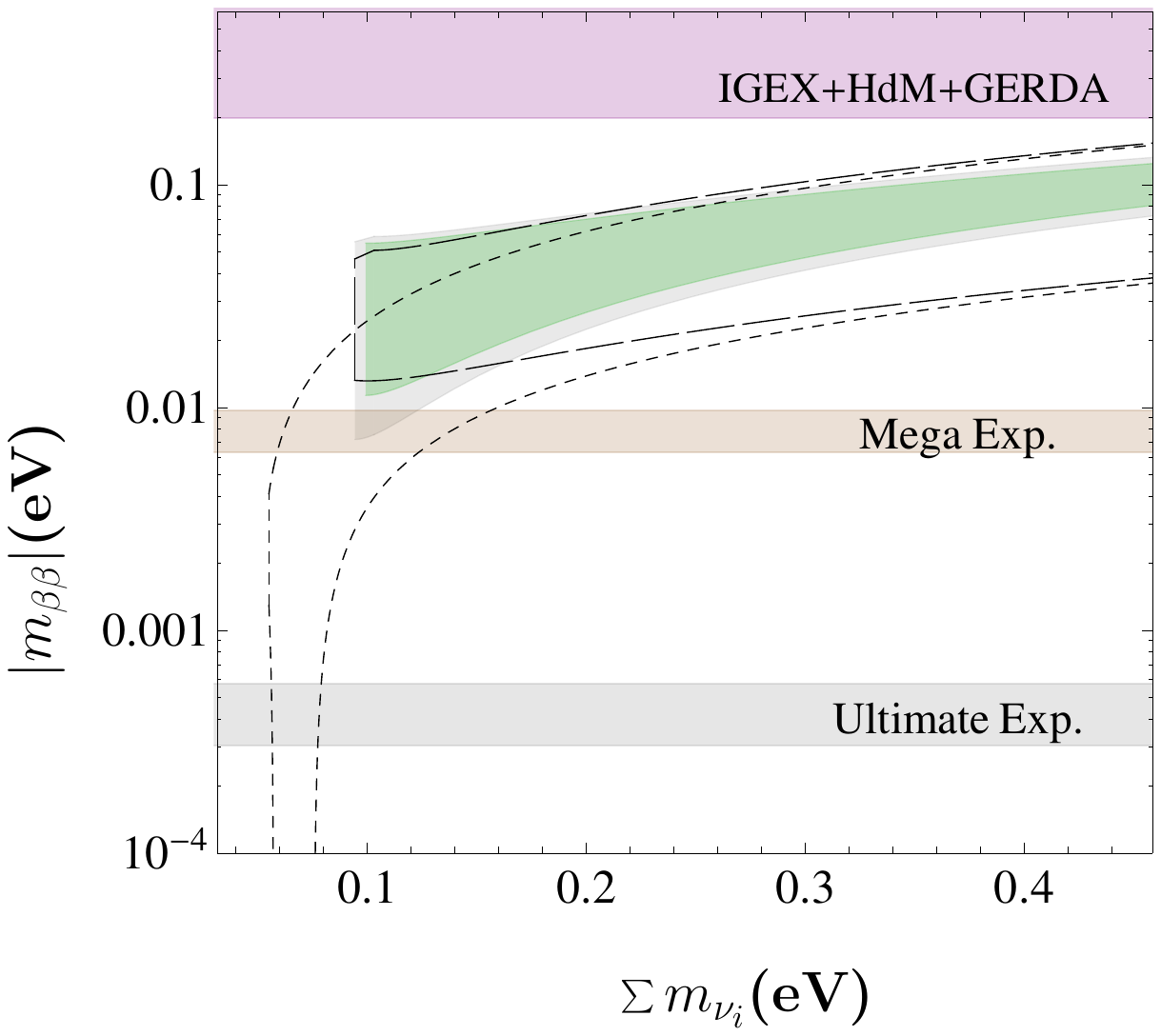}}
     \vspace*{-10pt}\caption{\label{fig:meffpseudosterile2IHapp}
      Plots for \meff vs $\sumass$ for the IH case in the presence of one extra sterile neutrino state and
    in the case in which respectively, from
    the left to the right,    $(\chi_1,\chi_2)$ or $(\chi_1,\chi_3)$ 
    or $(\chi_2,\chi_3)$ are Dirac fermions.  
     The light (Gray) and darker (Green) shaded areas are given as in Fig. \ref{fig:meffpseudosterileIHapp}. }
\end{figure}
\end{center}

 In this appendix for completeness we show the  \meff versus the sum of the three light 
 active states in the case one extra sterile neutrino is added, see Fig. \ref{fig:meffsterileapp}. 
 We show as well \meff vs the sum of the active neutrinos 
 in the pseudo-Dirac scenarios considered in Section \ref{sterile}.
 In Figs. \ref{fig:meffpseudosterileapp}, \ref{fig:meffpseudosterileIHapp},
 \ref{fig:meffpseudosterile2app} and \ref{fig:meffpseudosterile2IHapp} the light shaded regions correspond to the
3$\sigma$ uncertainty in the oscillation parameters while the 
darker shaded area determines the allowed range for the best fit values of ref. \cite{Capozzi:2013csa}.
\vspace*{-0cm}
\onecolumngrid
\vspace*{-1cm}
\vspace*{-1cm}
\vspace*{-1cm}
\vspace*{-1cm}

\bibliographystyle{h-physrev4.bst}
\bibliography{biblio}
\end{document}